\documentclass[structabstract]{aa}

\usepackage{txfonts}
\usepackage{natbib}
\usepackage{graphicx}
\usepackage{lscape}
\usepackage{longtable}

\begin{document}
\titlerunning{X-ray Cross-calibration with G21.5--0.9}
\authorrunning{M. Tsujimoto et al.}

\title{Cross-calibration of the X-ray Instruments\\ onboard the Chandra, INTEGRAL, RXTE,
Suzaku, Swift, and XMM-Newton Observatories using \object{G21.5--0.9}\thanks{This work
is based on the activity of the International Astronomical Consortium for High Energy
Calibration (IACHEC).}}
\subtitle{}
\author{
Masahiro Tsujimoto\inst{1}
\and
Matteo Guainazzi\inst{2}
\and
Paul~P. Plucinsky\inst{3}
\and
Andrew~P. Beardmore\inst{4}
\and
Manabu Ishida\inst{1}
\and
Lorenzo Natalucci\inst{5}
\and
Jennifer~L.~L. Posson-Brown\inst{3}
\and
Andrew~M. Read\inst{4}
\and
Richard~D. Saxton\inst{2}
\and
Nikolai~V. Shaposhnikov\inst{6}
}
\institute{
Japan Aerospace Exploration Agency, Institute of Space and Astronautical Science, 3-1-1
Yoshino-dai, Chuo-ku, Sagamihara, Kanagawa 252-5210, Japan
\and
European Space Agency, European Space Astronomy Centre, E-28691 Villanueva de la Ca\~{n}ada, Madrid, Spain
\and
Harvard-Smithsonian Center for Astrophysics, MS-70, 60 Garden Street, Cambridge, MA 02138, USA
\and
Department of Physics and Astronomy, University of Leicester, Leicester LE1 7RH, United Kingdom
\and
INAF, Istituto di Astrofisica Spaziale e Fisica Cosmica, Via del Fosso del Cavaliere, 100 00133 Roma, Italy
\and
National Aeronautics and Space Agency, Goddard Space Flight Center, Code 662, Laboratory for X-ray Astrophysics, Greenbelt, MD 20771, USA
}
\date{Received / Accepted }
\abstract
{The Crab nebula has been used as a celestial calibration source of the X-ray
flux and spectral shape for many years by X-ray astronomy missions. However, the object is often
too bright for current and future missions equipped with instruments with improved
sensitivity.}
{We use G21.5--0.9 as a viable, fainter substitute to the Crab, which is another
pulsar-wind nebula with a time-constant power-law spectrum with a flux of a few milli
Crab in the X-ray band. Using this source, we conduct a cross-calibration study of the
instruments onboard currently active observatories: Chandra ACIS, Suzaku XIS, Swift XRT,
XMM-Newton EPIC (MOS and pn) for the soft-band, and INTEGRAL IBIS-ISGRI, RXTE PCA, and
Suzaku HXD-PIN for the hard band.}
{We extract spectra from all the instruments and fit them under the same astrophysical
assumptions. We compare the spectral parameters of the G21.5--0.9 model: power-law
photon index, H-equivalent column density of the interstellar photoelectric absorption,
flux in the soft (2--8~keV) or hard (15--50~keV) energy band.}
{We identify the systematic differences in the best-fit parameter values unattributable
to the statistical scatter of the data alone. We interpret these differences as due to
residual cross-calibration problems. The differences can be as large as 20\% and 9\% for
the soft-band flux and power-law index, respectively, and 46\% for the hard-band
flux. The results are plotted and tabulated as a useful reference for future calibration
and scientific studies using multiple missions.}
{}
\maketitle

\section{INTRODUCTION}\label{s1}
From the successful launch to the end of the mission, all X-ray observatories put
tremendous effort into the flight calibrations of their instruments using celestial
objects to test their response models and to monitor changes in their performance. Most
such efforts have been made by each observatory independently from the others, using
similar methodologies and often the same targets. For the purpose of facilitating
communication and comparison of results among various calibration teams, an
international consortium for instrumental cross-calibration was established, which is
named the International Astronomical Consortium for High Energy Calibration
(IACHEC)\footnote{See http://www.iachec.org/index.html for detail.}.

As part of the IACHEC activities, we present the results of a comparison study of
G21.5--0.9 using the currently active X-ray missions. The purpose of this paper is to
compare the flux and the spectral shape in the X-ray bandpass of this time-constant
power-law source and to identify the systematic differences among various
instruments. Benefits for the community are two-fold: (1) The comparison result is
useful in improving the calibration and in interpreting data of other celestial objects
obtained by different instruments. (2) This study will be a basis to establish
G21.5--0.9 as a viable flight calibration source for the X-ray flux and the spectral
shape, substituting the traditionally-used Crab nebula \citep{kirsch05,weisskopf10},
which is often too bright for contemporary and future instruments with improved
sensitivity.

Participating missions for the comparison study are the Chandra X-ray Observatory
\citep{weisskopf02}, the International Gamma-ray Astrophysics Laboratory (INTEGRAL;
\citealt{winkler03}), the Rossi X-ray Timing Explorer (RXTE; \citealt{bradt93}), Suzaku
\citep{mitsuda07}, Swift \citep{gehrels04}, and XMM-Newton \citep{jansen01}
observatories.

We use the following instruments: the Advanced CCD Imaging Spectrometer (ACIS;
\citealt{garmire03}) onboard Chandra, the INTEGRAL Soft Gamma-Ray Imager (ISGRI;
\citealt{lebrun03}) equipped with the IBIS telescope \citep{ubertini03}, the
Proportional Counter Array (PCA; \citealt{jahoda96}) onboard the RXTE, the X-ray Imaging
Spectrometer (XIS; \citealt{koyama07}) and the Hard X-ray Detector (HXD;
\citealt{kokubun07,takahashi07}) PIN component onboard Suzaku, the X-Ray Telescope (XRT;
\citealt{burrows04}) onboard Swift, and the European Photon Imaging Camera (EPIC)
MOS-type \citep{turner01} and pn-type \citep{strueder01} CCDs onboard XMM-Newton. The
dispersive X-ray spectrometers onboard Chandra and XMM-Newton are out of the scope of
this study. Readers can also refer to \citet{snowden02}, in which two past missions ASCA
(GIS and SIS) and ROSAT (PSPC) are compared with Chandra (ACIS) and XMM-Newton (EPIC)
using the same object.

The instruments used in our study are divided into two groups: the soft-band instruments
(Chandra ACIS, Suzaku XIS, Swift XRT, XMM-Newton EPIC-MOS and EPIC-pn) sensitive below
$\sim$10~keV and the hard-band instruments (INTEGRAL IBIS-ISGRI, RXTE PCA, and Suzaku
HXD-PIN) sensitive above $\sim$10~keV. The instruments with imaging capability are all
the soft-band instruments with X-ray optics and INTEGRAL IBIS-ISGRI with coded mask
technique. All the soft-band instruments employ X-ray CCD devices, which have imaging and
medium-resolution ($E$/$\Delta E \sim$~30--50) spectroscopic capability.

\medskip

The plan of this paper is as follows. In \S~\ref{s2}, we describe the X-ray properties
of G21.5--0.9 and discuss the advantages and limitations to use this source as a
calibration source. In \S~\ref{s3}, we present the basic properties of the participating
instruments, the data sets, and the processing of the data. In \S~\ref{s4}, we first set
up the procedure for the event extraction and spectral fitting, which are common among
all the instruments as much as possible (\S~\ref{s4-1}). We then present the fitting
results of each individual instrument and justify the common procedure. We also discuss
the effect of some major systematic uncertainties arising from the procedure
(\S~\ref{s4-2}). We finally assemble all spectra for joint fitting (\S~\ref{s4-3}). In
\S~\ref{s5}, we compare the results among all the participating instruments and identify
systematic cross-calibration uncertainties. We also compare the differences found in the
present work with those in other studies. The paper concludes in \S~\ref{s6}.

\section{OBJECT --- G21.5--0.9 ---}\label{s2}
Pulsar wind nebulae (PWNe) are a small class of supernova remnants with intrinsically
bright non-thermal X-ray emission, which includes the prototypical Crab nebula, 3C\,58,
Kes\,75, PSR B0540--69, PSR B1509--58, and G21.5--0.9. PWNe are suited for X-ray flux
calibrations for their constant total X-ray flux over human time scales. The
brightness and the power-law index of the spectrum change spatially across a PWN, but
the integrated spectrum can be represented by a single power-law in practice. Therefore,
they can also be used for calibrating the spectral shape, i.e., the power-law index in
the X-ray photon spectra. Indeed, the Crab nebula has been used as a standard X-ray flux
and spectral shape calibration source for decades.

G21.5--0.9 is a PWN powered by the pulsar PSR\,J1833--1034 with a period of 61.8~ms
\citep{gupta05,camilo06}. The age and distance of the object are estimated as 870~yr
\citep{bietenholz08} and 4.8~kpc \citep{tian08}, respectively. The object is a bright
X-ray emitter. The Einstein observatory resolved the object into several structures of
different spatial scales \citep{becker81}. The EXOSAT and Ginga satellites showed that
the X-ray spectrum is represented by a power-law model \citep{davelaar96,asaoka90},
arguing for the PWN nature of this source.

G21.5--0.9 has several advantages over other PWNe as a celestial calibration source: (1)
It is fainter than the Crab nebula by a factor of $\sim$500, which places it in a
reasonable flux range to match with the dynamic ranges of contemporary and future
missions. (2) It is compact in size for its distance and youth, mitigating the effects
caused by spatial differences of mirror and detector responses. (3) It has a very simple
circular distribution of brightness and spectral hardness presumably because of the
pole-on geometry, making the source and background extraction easy. (4) The spectrum is
flatter than other PWNe, thus the object can be used to calibrate detectors over a wide
range of energies including those above 10~keV. (5) The soft ($<$1~keV) photons
contribute little to the spectrum due to a large interstellar extinction of
$\gtrsim$10$^{22}$~cm$^{-2}$. Therefore, the calibration uncertainty stemming from the
accumulation of contaminating material on CCDs, which plagues both ACIS and XIS, can be
decoupled. (6) No pulsation is confirmed in the X-ray band, which indicates that the
spectrum does not change in a pulse phase unlike the Crab nebula. (7) Because of these
properties, G21.5--0.9 has been used as a calibration source for Chandra, Swift, and
XMM-Newton. We can exploit the wealth of existing data set.

The source also has some limitations: (1) G21.5--0.9 is extended by $\sim$3\arcmin\ and
is not adequate to calibrate dispersive spectrometers. (2) The object is located close
to the Galactic plane with a rich population of bright X-ray binaries, which makes
source confusion non-negligible for instruments without imaging capability (PCA and
HXD-PIN). (3) The spectrum suffers a large interstellar extinction, leaving few photons
in the energy band below $\sim$1~keV. Soft-band calibration sources need to be
complemented, which is covered by other IACHEC studies (e.g.,
\citealt{beuermann06,plucinsky08}).

\begin{figure}[hbtp]
 \begin{center}
  \resizebox{\hsize}{!}{\includegraphics{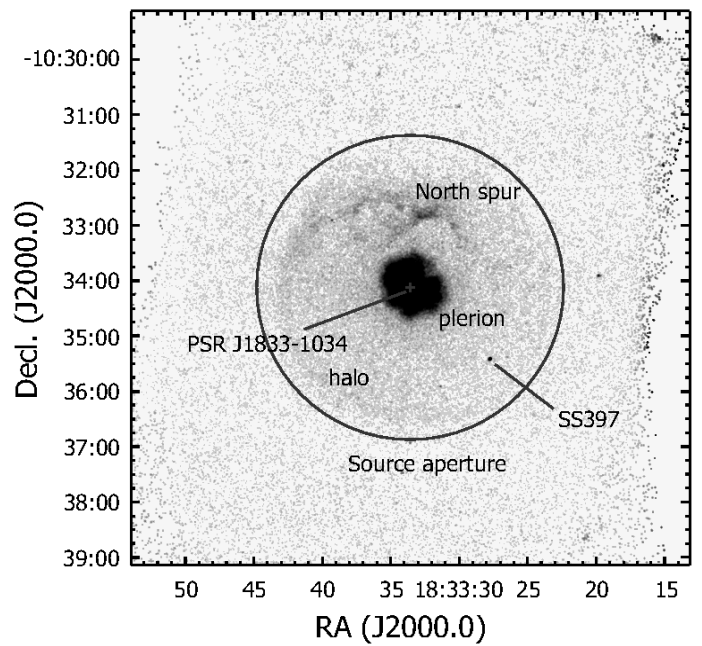}}
 \end{center}
 \caption{Chandra ACIS-S3 image of G21.5--0.9. The spatial components and the nearby
 source SS\,397 are labeled. The source extraction region of 165\arcsec\ is indicated by
 the solid circle.}
 \label{f1}
\end{figure}

Several spatial structures are known in the X-ray image. Figure~\ref{f1} shows a Chandra
ACIS-S3 view of the object. At the center, the ``core'' is found \citep{slane00}, which
is the neutron star with a power-law spectrum. The X-ray pulsation is yet to be found
from the pulsar \citep{lapalombara02,camilo06}. The core is actually resolved with a
size of $\sim$2\arcsec, possibly representing the pulsar wind termination shock
\citep{slane00}. Surrounding this is the ``plerion'' with $\sim$30\arcsec\ in size,
which is a synchrotron nebula. The spectral hardness decreases as the distance from the
core increases, indicating the synchrotron loss of high energy electrons as a function
of energy. An even larger but much fainter structure is found around the plerion, which
is called ``halo'' with $\sim$150\arcsec\ in size. The halo also shows a power-law
spectrum with decreasing hardness as distance increases \citep{warwick01}. The origin of
the halo is different from that of the plerion. One idea is that the halo is the dust
scattered light from the plerion \citep{bandiera04}. Upon these non-thermal features, a
faint knot is found $\sim$1\farcm5 north of the core, which is called the ``North
Spur''. Based on the thermal spectral features, it is considered to be ejecta from the
supernova explosion \citep{bocchino05}.

Some PWNe (e.g., the Crab nebula) show time-variable filaments. In G21.5--0.9,
\citep{matheson10} have used archival Chandra data to detect some localized variable
emission near the pulsar. However, the contribution of these variable features to the
total flux of G21.5-0.9 is negligible of $\lesssim$1\% and therefore our analysis is not
affected by these features.

An unrelated point source is found $\sim$2\farcm0 southwest of the core. The object is
identified as an emission-line star SS\,397 \citep{stephenson77,warwick01}. The source
is a $\gamma$ Cas analogue, which is a small class of B\textit{e} stars known to show
weakly variable X-ray flux with extreme hardness.

All these spatial structures can be separated only with telescopes with a good spatial
resolution. Because the participating instruments in our cross-calibration study include
those with low spatial resolution or the lack thereof, we use a large enough source
extraction region to encompass all of these structures. We ignore the contaminating
thermal emission from the North Spur, which is very faint at a level of $\sim$0.1\% of
the non-thermal emission. We retrieved all the Chandra archive to examine the behavior
of the unrelated neighboring source SS\,397, and found that the source is variable by a
factor of a few, but has a negligible flux even at the brightest level. The X-ray
spectrum is dominated by the integration of spatially-varying power-law emission. We
only treat the integrated spectrum, which we approximate as a single power-law
spectrum. The reduced $\chi^{2}$ values in our fitting will show that this approach is
valid for the present data set.

\section{INSTRUMENTS, DATA, \& PROCESSING}\label{s3}
\begin{table*}
 \caption{List of data sets and their basic properties.}\label{t1}
 \centering
  \begin{tabular}{cccccrccccc}
   \hline 
   Label & Obser- & Instru- & ObsID & Date & $t_{\rm{exp}}$\tablefootmark{a} & Band\tablefootmark{b} &
   $C_{\rm{src}}$\tablefootmark{c} & Cnt\tablefootmark{d} & $r_{\rm{in}}$\tablefootmark{e} & $r_{\rm{out}}$\tablefootmark{e} \\
    & vatory & ment &  &  & (ks) & (keV) & & bin$^{-1}$ & (\arcmin) & (\arcmin) \\
   \hline
   CS0 & Chandra & ACIS-S3 & 1717 & 2000-05-23 & 7.5 & 1.0--8.0 & 28372 & 50 & ... & ... \\
   CS1 &         &         & 2873 & 2002-09-14 & 9.8 & 1.0--8.0 & 36370 & 50 & ... & ... \\
   CS2 &         &         & 3700 & 2003-11-09 & 9.5 & 1.0--8.0 & 34749 & 50 & ... & ... \\
   CS3 &         &         & 5166 & 2004-03-14 & 10  & 1.0--8.0 & 35777 & 50 & ... & ... \\
   CS4 &         &         & 5159 & 2004-10-27 & 9.8 & 1.0--8.0 & 36729 & 50 & ... & ... \\
   CS5 &         &         & 6071 & 2005-02-26 & 9.8 & 1.0--8.0 & 34968 & 50 & ... & ... \\
   CS6 &         &         & 6741 & 2006-02-22 & 9.8 & 1.0--8.0 & 35947 & 50 & ... & ... \\
   \hline
   CS7 &         &         & 1553 & 2001-03-18 & 9.7 & 1.0--8.0 & ... & 50 & ... & ... \\
   CS8 &         &         & 1554 & 2001-07-21 & 9.1 & 1.0--8.0 & ... & 50 & ... & ... \\
   CS9 &         &         & 3693 & 2006-05-16 & 9.8 & 1.0--8.0 & ... & 50 & ... & ... \\
   \hline 
   IS0 & INTEGRAL & ISGRI & ... & 2003--2008 & 3132 & 18--150 & 3.39$\times$10$^{6}$ & ... & ... & ... \\
   \hline 
   RP0 & RXTE     & PCA   & 20259-01-01-000  & 1997-11-08 & 21 & 5.0--30 & 1.27$\times$10$^{6}$ &  ... & ... & ... \\
   \hline
   SI0 & Suzaku & XIS0 & 104023010 & 2009-10-10 & 40 & 1.0--8.0 & 81297 & 100 & 5.0 & 7.0 \\
   SI1 &        & XIS1 &           & 2009-10-10 & 40 & 1.0--8.0 & 89436 & 100 & 5.0 & 7.0 \\
   SI3 &        & XIS3 &           & 2009-10-10 & 40 & 1.0--8.0 & 83310 & 100 & 5.0 & 7.0 \\
   SP0 &        & PIN  &           & 2009-10-10 & 30 & 15--70 & 12684 & 800 & ... & ... \\
   \hline
   SX0 & Swift & XRT & 00053600001 & 2006-08-13 & 17 & 1.0--8.0 & 10593 & 20 & 5.0 & 7.0 \\
       &       &     & 00053600002 & 2006-08-15 &    &          & & & & \\
       &       &     & 00053601001 & 2006-08-23 &    &          & & & & \\
       &       &     & 00053601002 & 2006-08-24 &    &          & & & & \\
   SX1 &       &     & 00053600004 & 2007-05-09 & 26 & 1.0--8.0 & 15767 & 20 & 5.0 & 7.0 \\
       &       &     & 00053600006 & 2007-05-11 &    &          & & & & \\
       &       &     & 00053600007 & 2007-05-16 &    &          & & & & \\
       &       &     & 00053600008 & 2007-05-17 &    &          & & & & \\
       &       &     & 00053600009 & 2007-05-29 &    &          & & & & \\
       &       &     & 00053600010 & 2007-05-31 &    &          & & & & \\
       &       &     & 00053600011 & 2007-07-04 &    &          & & & & \\
       &       &     & 00053600012 & 2007-06-28 &    &          & & & & \\
   \hline
   SX2 &       &     & 00053600021 & 2007-10-06 & 28 & 1.0--8.0 & 17999 & 20 & 5.0 & 7.0 \\
       &       &     & 00053600025 & 2007-10-12 &    &          & & & & \\
       &       &     & 00053600031 & 2007-10-24 &    &          & & & & \\
       &       &     & 00053600032 & 2007-10-25 &    &          & & & & \\
   SX3 &       &     & 00053600033 & 2009-10-16 & 27 & 1.0--8.0 & 16392 & 20 & 5.0 & 7.0 \\
       &       &     & 00053600034 & 2009-10-18 &    &          & & & & \\
   \hline
   XM1 & XMM  & MOS1 & 0122700101 & 2000-04-07 & 29 & 1.0--8.0 & 77541 & 200 & 3.3 & 5.0 \\
   XM2 &      & MOS2 &            & 2000-04-07 & 29 & 1.0--8.0 & 76310 & 200 & 3.3 & 5.0 \\
   XP0 &      & pn   &            & 2000-04-07 & 24 & 1.0--8.0 & 172356 & 200 & 0.0 & 2.75 \\
   \hline
  \end{tabular}
  \tablefoot{
  \tablefoottext{a}{Net exposure time after cleaning the events.}
  \tablefoottext{b}{Energy range used in the spectral fitting.}
  \tablefoottext{c}{Number of counts in the source extraction region in the fitting energy range.}
  \tablefoottext{d}{Number of counts per spectral bin for the fitting.}
  \tablefoottext{e}{Inner and outer radii for the background extraction annulus. For
 ACIS-S3 data, we used a rectangular region (\S~\ref{s4-2-1}). For EPIC-pn, we used
 blank sky data (\S~\ref{s4-2-6}).}
  }
\end{table*}

\subsection{Chandra ACIS}
\subsubsection{Instruments}
The ACIS is an X-ray imaging-spectrometer consisting of the ACIS-I and ACIS-S CCD
arrays. The imaging capability is unprecedented with a half-power diameter (HPD) of
$\sim$1\arcsec\ at the on-axis position. We use one of the CCD chips (ACIS-S3) in 
the ACIS-S array. The ACIS-S3 chip is back-side illuminated device sensitive at a
0.2--10~keV band. The chip has 1024$\times$1024 pixels covering a
8\farcm4$\times$8\farcm4 area.

\subsubsection{Data}
\begin{figure}[hbtp]
 \begin{center}
  \resizebox{\hsize}{!}{\includegraphics{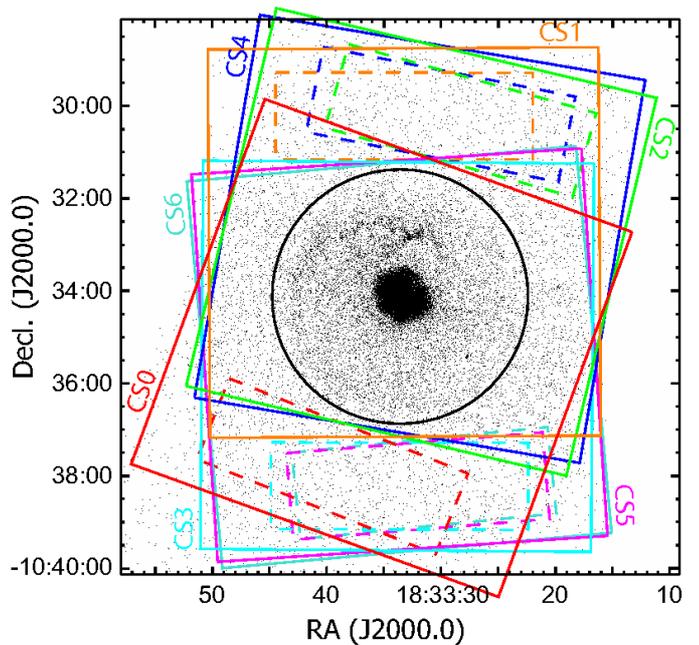}}
 \end{center}
 \caption{Layout of the field of views (solid squares) and the background extraction
 regions (dashed rectangles) for each ACIS-S3 full-frame data set (CS0--6) in different
 colors. The source extraction region (solid circle) is displaced from the image center,
 leaving the other side of the image for the background extraction of a
 5\farcm5$\times$1\farcm9 region.}
 \label{f2}
\end{figure}

The Chandra data were retrieved from the archive. A total of 41 ACIS observations were
available with G21.5-0.9 at various detector positions. We chose seven full-frame
observations with the object close to but offset from the center of the ACIS-S3 chip
(CS0--CS6; table~\ref{t1}), so that the chip encompasses the entire spatial structure of
the source and leaves some area for background event extraction (figure~\ref{f2}). We
also supplemented three ACIS-S3 sub-array data (CS7--CS9; table~\ref{t1}) to evaluate
the effect of pile-up (\S~\ref{s4-2-1}). The data were taken in the timed exposure mode
with a frame time of 3.2~s for the full frame (CS0--CS6) observations or of 0.84~s for
the sub-array (CS7--CS9) observations.

\subsubsection{Processing}
We processed the data based on the calibration database (CALDB) version 4.2 using the
CIAO package version 4.2. The redistribution matrix functions (RMFs) and the auxiliary
response files (ARFs) were generated for each data set. The RMFs take into account the
spatial dependence of the detector response. The ARFs take into account the flux outside
of the extraction region, although this is not a significant issue for the ACIS-S3 data
as the resolved object is entirely contained in the source extraction region
(figure~\ref{f2}).

\subsection{INTEGRAL IBIS-ISGRI}
\subsubsection{Instruments}
The IBIS telescope is a payload of INTEGRAL. It is composed of two layers of
detectors. The upper layer is ISGRI sensitive in an energy range of 15--1000~keV
band. ISGRI is equipped with a CdTe detector array of 128$\times$128 pixels with a
spectral resolution of $\sim$9\% at 100 keV. The IBIS has a coded mask, giving the
instrument imaging capability of an unprecedented angular resolution of 12\arcmin\ in
FWHM in this band. The full-coded field of view is $\sim$9\degr\ square.

\subsubsection{Data}
The IBIS data were obtained from a set of $\sim$4000 IBIS Science Window data sets taken
in a time period of 5 years from March 2003 to March 2008. We checked the source light
curve in the 18--60~keV band and found that it is constant at a count rate of
$\sim$0.9~s$^{-1}$. We thus stacked all the data sets.

\subsubsection{Processing}
The standard INTEGRAL Off-line Scientific Analysis (OSA) version 8.0 was used to obtain
images and spectra. We used the standard IBIS response files released in the OSA 8.0
distribution. The IBIS ARFs are computed in each year to incorporate the updated
systematic uncertainties, which could affect the analysis of bright sources. Due to the
very long time span for the observations of this source, we computed an averaged ARF for
the total spectrum, which was weighted by the total exposure time in each calibration
period.

\subsection{RXTE PCA}
\subsubsection{Instruments}
The PCA is a proportional counter array consisting of five proportional counter units
(PCU1--5). The counters are filled with Xe gas and are sensitive at 2--60~keV. The
instrument is unique in its high timing accuracy, a large effective area, and modest
spectral resolution of $\sim$1~keV at 6~keV. There are no focussing optics but the
collimator limits the field of view to a $\sim$1\degr\ square.

\subsubsection{Data}
We retrieved the archived data of G21.5-0.9 observed on 1997 November 8
(table~\ref{t1}). We extracted the PCA spectrum from the standard2 PCA data mode, which
provides spectral information in 129 channels with a time resolution of 16~s.

The electron flux stayed lower than 0.2 counts~s$^{-1}$~PCU$^{-1}$ and no PCU breakdown
events occurred during the entire observation. Data from all PCUs and all layers were
combined into one spectrum. All five PCUs were operational during the entire
observation, except for South Atlantic Anomaly (SAA) passages.

\subsubsection{Processing}
We excluded events when the elevation angle from the Earth rim was smaller than 10$^\circ$
and during SAA passages and 30 minutes thereafter. As a result, we obtained a net
exposure time of 21~ks.

The background spectrum and the RMF were calculated using the latest versions of
\texttt{pcabackest} (version 3.8) and \texttt{pcarmf} (version 11.7) tools
respectively. The background spectrum was scaled down by 1.3\% to match the Standard2
rate in channels 110--128, corresponding to energies higher than 75~keV where no source
counts are expected.

\subsection{Suzaku XIS and HXD-PIN}
\subsubsection{Instruments}
The XIS is an X-ray imaging-spectrometer equipped with four X-ray CCDs sensitive in the
0.2--12~keV band. One is a back-side illuminated (XIS1) device and the others are front-side
illuminated (XIS0, 2, and 3) devices. The entire XIS2 and a part of the XIS0 are
nonfunctional due to putative micro-meteorite hits, hence are not used in this
paper. The four CCDs are located at the focal plane of four co-aligned X-ray telescopes
with a HPD of $\sim$2\farcm0. Each XIS sensor has 1024$\times$1024 pixels and covers a
17\farcm8$\times$17\farcm8 view, which encompasses the entire G21.5--0.9 structure and
leaves ample room to construct background spectra.

The PIN is a component of the HXD covering 10--70~keV. It is a non-imaging detector
composed of 64 Si PIN diodes at the bottom of well-type collimators surrounded by GSO
anti-coincidence scintillators. The effective area monotonically decreases as the
distance increases from the field center, with a full width at zero intensity (FWZI)
view of $\sim$70\arcmin\ square.

\subsubsection{Data}
We executed a Suzaku observation dedicated for the present study on 2009 October 10 for
40~ks using the XIS and the HXD. The target was placed at the XIS nominal position. The
XIS was operated in the normal clocking mode with a frame time of 8~s.

Within the FWZI square view of the PIN, two other sources were found in the INTEGRAL
source catalogue \citep{bird10}; 4U\,1835--11 and AX\,J1831.2--1008 respectively at
26\farcm4 and 42\farcm7 from the field center. Because both of the unrelated sources
have intensities less than 0.2\% of G21.5--0.9 in the ISGRI 20--60~keV band, we ignore
the contribution of these sources in the PIN spectrum.

\subsubsection{Processing}
The data were processed through the pipeline processing version 2.4 and were reduced
using the HEASoft version 6.8--6.9.

For the XIS, the RMFs and ARFs were generated using the \texttt{xisrmfgen} and
\texttt{xissimarfgen} tools, respectively. The former takes into account the
positional difference of the detector response. The latter is based on a ray-tracing
simulation, which is designed to compensate for the lost photons outside of the photon
extraction region. The tool also takes into account the surface brightness distribution
of incoming X-ray sources to calculate the effective area with the assumption that the
spectrum is uniform across the emitting region. We used the combined ACIS-S3 image
(figure~\ref{f1}) as the true spatial distribution.

For the PIN, the background consists of astrophysical and instrumental background. The
former is dominated by the cosmic X-ray background (CXB). We used the instrumental
background spectrum and the detector response files distributed by the instrument
team. We simulated the CXB spectrum using the energy response and added it to the
instrumental background to subtract the CXB contribution. G21.5--0.9 is assumed to be
point-like to generate the spatial response of the detector.

\subsection{Swift XRT}
\subsubsection{Instruments}
The XRT comprises a Wolter-I telescope, which focuses X-rays onto an X-ray CCD device
identical to the ones flown on the XMM-Newton EPIC-MOS instrument \citep{burrows04}. The
CCD, which is responsive to 0.2--10~keV X-rays, has a dimension of 600$\times$600
pixels, covering a 23\farcm6$\times$23\farcm6 field. The mirror has a HPD of
$\sim$18\arcsec\ and gives astrometric accuracies of a few arcseconds.

As Swift's primary science goal is to rapidly respond to gamma-ray bursts (GRBs), the
XRT was designed to operate autonomously, so that it could measure GRB light curves and
spectra over seven orders of magnitude in flux. In order to mitigate the effects of
pile-up, the XRT automatically switches between different CCD readout modes depending
on the source brightness. Two frequently-used modes are: Windowed Timing (WT) mode,
which provides 1D spatial information in the central 7\farcm8 of the CCD with a time
resolution of 1.8 ms, and Photon Counting (PC) mode, which allows full 2D
imaging-spectroscopy with a time resolution of 2.5~s.

The XRT's thermoelectric cooler power supply system apparently failed shortly after
launch, resulting in CCD operating temperatures ranging from --75 to --50 $^{\circ}$C
instead of the intended nominal --100 $^{\circ}$C. In order to reduce the dark current
and overall noise level caused by the higher operating temperatures, the CCD substrate
voltage ($V_{\rm{ss}}$) was raised from 0 to 6~V on 2007 August 30. While this allowed
the CCD to operate a few degrees warmer, it had the minor drawback of reducing the
quantum efficiency (QE) both at high energies ($\gtrsim$5~keV) and just below the Si
edge (1.5--1.84~keV). The $V_{\rm{ss}}=$ 6~V calibration modifications required to
correct the changes of QE have been completed for the WT mode but not for the PC mode,
at the time of writing; the latter will be released in due course.

\subsubsection{Data}
G21.5--0.9 is used as a routine calibration source for the XRT as its heavily absorbed
spectrum is useful to verify the high energy redistribution properties of the CCD. As the
object is an extended source and it is difficult to subtract background adequately for
the WT data, we present only results from PC mode observations.

As Swift has a flexible observing schedule, often interrupted by GRBs, observations are
divided into ObsIDs containing one or more snapshots, where each snapshot has a typical
exposure of 1--2~ks. For a faint source like G21.5--0.9, it is necessary to accumulate
data from different ObsIDs for any given epoch.

We retrieved 18 observations from the UK Swift Science Data Centre. The observations
were selected so that the source lay within 10\arcmin\ of the CCD bore-sight in order to
avoid pointings that have large off-axis angles to minimize the effect of
vignetting. The data were grouped into four epochs, split into two before (SX0 and SX1)
and after (SX2 and SX3) the $V_{\rm{ss}}$ change (table~\ref{t1}). Typical exposures
were 17--18~ks per epoch.

\subsubsection{Processing}
The data were processed with the Swift software version 3.5, which was made available
with HEASoft version 6.8.

For each epoch, a summed exposure map was computed, which accounts for the CCD
bad columns and the telescope vignetting. The exposure maps were then used to create
extended source ARF files, by subdividing the source extraction region into boxes and
calculating an average ARF over the regions, weighted by the relative exposure and
counts in each box. For this analysis, the CALDB v011 RMF and ARFs were used.

\subsection{XMM-Newton EPIC}
\subsubsection{Instruments}
The EPIC is composed of two different types of X-ray CCD devices; two units of MOS-type
(MOS1 and MOS2) covering 0.15--12~keV and one unit of pn-type covering 0.15-15 keV. The
three devices are located at the focal plane of three co-aligned telescopes with a HPD
of $\sim$15\arcsec.

The MOS1 and MOS2 cameras consist of an array of seven front-illuminated CCD chips, each
of which has 600$\times$600 pixels covering a $\sim$10\farcm9$\times$10\farcm9 square
region. The pn-type has a 2$\times$6 back-illuminated CCD array, each of which has
200$\times$64 pixels covering a $\sim$13\farcm6$\times$4\farcm4 rectangular region.

\subsubsection{Data}
XMM-Newton observed G21.5--0.9 several times within the field of view of the EPIC
camera. We used the data with the target placed at the center of the camera
taken on 2000 April 7 (table~\ref{t1}). The observation was performed with a
medium-thickness filter and a full-frame mode for the CCD clocking.

\begin{figure}[hbtp]
 \begin{center}
  \resizebox{\hsize}{!}{\includegraphics{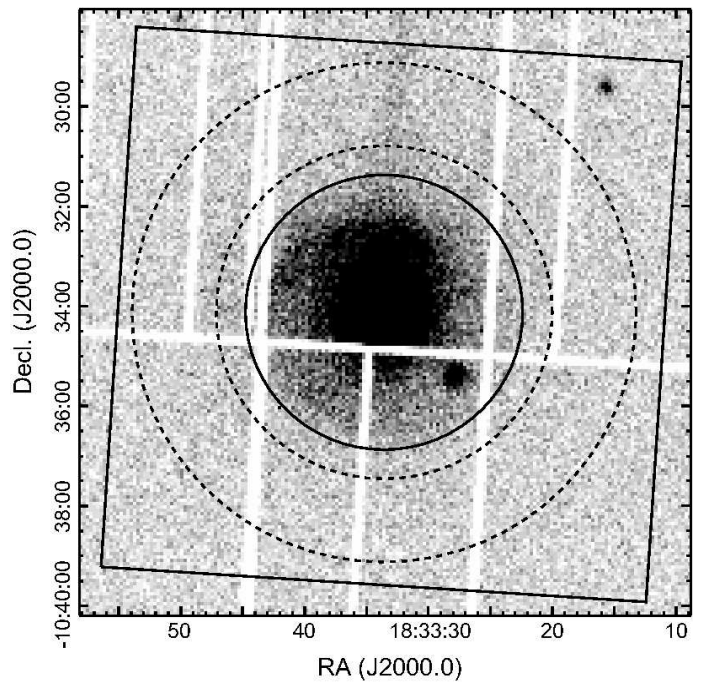}}
 \end{center}
 \caption{Layout of the field of view of the central chip of the two MOS arrays (solid
 square), the MOS and pn source extraction region (solid circle), and the MOS background
 extraction region (dashed annulus) overlaid on the pn image. The pn chip gaps and bad
 columns are apparent in the image.}
\label{f3}
\end{figure}

In the MOS arrays, the object is located at the center of the central CCD of both
arrays. The entire spatial structure is contained in a chip. In the pn-array, G21.5--0.9
is located at the bottom of CCD number 4 and the structure spreads across neighboring
CCDs (figure~\ref{f3}).

\subsubsection{Processing}
The data were reduced using the XMM-Newton Science Analysis System (SAS) version 10
\citep{gabriel04} using the most updated calibration available at the time of reducing
the data (August 2010). Event lists were generated using the data reduction meta-tasks
\texttt{e[mp]proc} with the default settings. The data were screened to remove intervals
of high particle background.

Response files were generated using the \texttt{rmfgen} and \texttt{arfgen} tasks. We
took into account the spatial distribution of the extended X-ray emission by calculating
a coarse detector map around the G21.5-0.9 centroid, which was convolved with the
spatially-dependent effective area to produce the appropriate transfer function for the
extraction region. The SAS does not support the correction for the photons lost outside
of the source extraction region. However, we consider that this is negligible by taking
a source extraction region large enough to encompass most photons to the level of
10$^{-3}$ of the peak (figure~\ref{f4}).

\section{ANALYSIS \& RESULTS}\label{s4}
\subsection{Overall procedure}\label{s4-1}
\subsubsection{Source Extractions for Soft-band Instruments}\label{s4-1-1}
\begin{figure}[hbtp]
 \begin{center}
  \resizebox{\hsize}{!}{\includegraphics{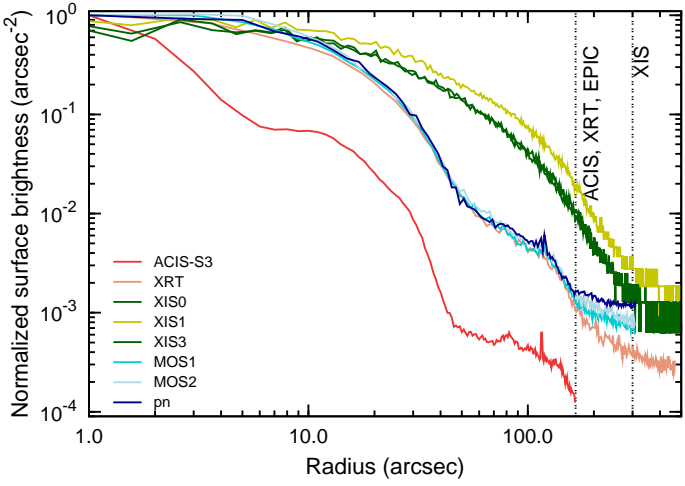}}
 \end{center}
 \caption{Radial profile of background-unsubtracted 0.5--8~keV counts for the
 soft-band instruments. The counts are normalized to the value at the center. The radii
 of the source extraction apertures are shown with the dashed lines.}
\label{f4}
\end{figure}

For the soft-band instruments, we extracted source events from a 165\arcsec\ circle
centered at the pulsar PSR\,J1833--1034 at (RA, Dec) $=$ (18:33:33.57,-10:34:07.5) in
the equinox J2000.0. The radius was chosen as a compromise (1) to be large enough to
encompass a significant fraction of the entire structures of G21.5--0.9 as well as the
negligible unrelated emission (figure~\ref{f4}) and (2) to be small enough to be
contained in the field of view of all the instruments. The Suzaku XIS with a larger
telescope HPD than the others is the only exception; we chose a larger source extraction
radius of 300\arcsec, so that it contains a significant fraction of the source photons
(figure~\ref{f4}). The background events were extracted from various regions optimized
for each instrument, which is discussed individually in \S~\ref{s4-2}.

\subsubsection{Fitting Procedure}\label{s4-1-2}
We assembled the source and the background spectra, RMFs, and ARFs for all the
instruments. Using the \texttt{XSPEC} package version 12.6, we fitted the
background-subtracted spectra. We applied the spectral model \texttt{pegpwrlw}
attenuated by the interstellar photoelectric absorption model \texttt{TBabs}
\citep{wilms00}. The former is the same with the conventional power-law model except
that it uses the flux between the designated lower and upper energy as a free parameter
for the intensity, not the normalized flux at 1~keV. This decouples the often entangled
parameters of the index of power ($\Gamma$) and the normalization in power-law
fitting. For the soft-band and hard-band instruments, the flux in the energy range of
2--8~keV ($F_{\rm{X,soft}}$) and 15--50~keV($F_{\rm{X,hard}}$) was evaluated,
respectively. \citet{verner96} was used for the photoelectric absorption cross section,
while \citet{wilms00} was used for the metal abundance for the derivation of the
H-equivalent extinction column density ($N_{\rm{H}}$).

For the soft-band instruments, we used the common energy range of 1--8~keV for the
fitting. All are Si-based CCDs, and thus have sensitivity in a similar energy range.
For the hard-band instruments, on the other hand, we used different ranges optimized for
each instrument: 18--150~keV for INTEGRAL ISGRI, 5.0--30~keV for RXTE PCA, and
15--70~keV for Suzaku PIN. The inhomogeneity of the detecting technique did not allow us
to use a common energy range. Because the power-law index of PWN spectra changes with
the energy range, the choice of different fitting energy range might introduce an
additional systematic uncertainty.

Spectra were binned with the number of counts per bin specified in table~\ref{t1}. The
model was fitted to minimize the $\chi^{2}$ statistics and was considered statistically
acceptable if the null hypothesis probability was larger than 5\%. Statistical
uncertainties for the best-fit values were derived as the 1 $\sigma$ deviations.

\begin{table*}
 \caption{Best-fit parameters and uncertainties in the spectral fitting.\tablefootmark{a}}\label{t2}
 \centering
  \begin{tabular}{cccccc}
   \hline 
   Label & $N_{\rm{H}}$\tablefootmark{b} & $\Gamma$\tablefootmark{c} &
   $F_{\rm{X,soft}}$\tablefootmark{d} & $F_{\rm{X,hard}}$\tablefootmark{e} &
   Red-$\chi^{2}$\tablefootmark{f} \\
   & (10$^{22}~$cm$^{-2}$) & & \multicolumn{2}{c}{(10$^{-11}$~erg~s$^{-1}$~cm$^{-2}$)}
	       & /d.o.f.\\
   \hline
   \multicolumn{6}{c}{Chandra ACIS-S3} \\
   \hline
   CS0 & 2.99 (2.93--3.04) & 1.83 (1.80--1.86) & 6.10 (6.05--6.16) & ... & 0.93/ 302\\
   CS1 & 3.07 (3.01--3.12) & 1.85 (1.83--1.88) & 6.09 (6.04--6.13) & ... & 0.90/ 326\\
   CS2 & 3.04 (2.98--3.09) & 1.82 (1.79--1.84) & 6.06 (6.01--6.11) & ... & 1.04/ 325\\
   CS3 & 3.11 (3.05--3.16) & 1.84 (1.81--1.87) & 6.04 (5.99--6.09) & ... & 0.89/ 327\\
   CS4 & 3.16 (3.11--3.22) & 1.88 (1.85--1.91) & 6.10 (6.05--6.15) & ... & 1.03/ 330\\
   CS5 & 3.00 (2.95--3.06) & 1.81 (1.78--1.84) & 6.01 (5.97--6.06) & ... & 1.06/ 327\\
   CS6 & 3.14 (3.08--3.20) & 1.88 (1.85--1.91) & 6.03 (5.98--6.08) & ... & 1.07/ 326\\
   CS0--6 & 3.07 (3.05--3.09) & 1.84 (1.83--1.85) & 6.06 (6.04--6.08) & ... & 0.99/2281\\
   \hline
   \multicolumn{6}{c}{INTEGRAL IBIS-ISGRI} \\
   \hline
   IS0 & 2.99 & 2.09 (2.02--2.16) & ... & 4.17 (4.03--4.31) & 1.43/   9\\
   \hline
   \multicolumn{6}{c}{RXTE PCA} \\
   \hline
   RP0 & 2.99 & 2.05 (2.04--2.07) & ... & 5.54 (5.45--5.64) & 0.61/  51\\
   \hline
   \multicolumn{6}{c}{Suzaku XIS} \\
   \hline
   SI0 & 3.07 (3.03--3.11) & 1.91 (1.89--1.92) & 5.62 (5.59--5.65) & ... & 1.04/ 616\\
   SI1 & 3.18 (3.14--3.22) & 1.92 (1.90--1.93) & 5.73 (5.70--5.76) & ... & 1.06/ 632\\
   SI3 & 3.07 (3.03--3.11) & 1.90 (1.88--1.92) & 5.66 (5.63--5.69) & ... & 0.98/ 624\\
   \hline
   \multicolumn{6}{c}{Suzaku HXD-PIN} \\
   \hline
   SP0 & 2.99 & 2.28 (2.14--2.42) & ... & 6.10 (5.79--6.42) & 1.40/  12\\
   \hline
   \multicolumn{6}{c}{Swift XRT} \\
   \hline
   SX0 & 2.97 (2.88--3.07) & 1.77 (1.73--1.81) & 5.79 (5.72--5.87) & ... & 0.99/ 421\\
   SX1 & 2.90 (2.83--2.98) & 1.77 (1.74--1.81) & 5.48 (5.42--5.54) & ... & 1.03/ 479\\
   SX2 & 3.05 (2.98--3.13) & 1.90 (1.87--1.94) & 5.46 (5.40--5.51) & ... & 1.07/ 488\\
   SX3 & 3.16 (3.08--3.25) & 1.93 (1.89--1.96) & 5.46 (5.40--5.52) & ... & 1.14/ 478\\
   SX0$+$1 & 2.93 (2.87--2.99) & 1.77 (1.75--1.80) & 5.61 (5.56--5.65) & ... & 1.02/ 903\\
   SX2$+$3 & 3.10 (3.05--3.16) & 1.91 (1.89--1.94) & 5.46 (5.41--5.50) & ... & 1.11/ 969\\
   \hline
   \multicolumn{6}{c}{XMM-Newton EPIC} \\
   \hline
   EM1 & 2.90 (2.87--2.94) & 1.78 (1.77--1.80) & 5.47 (5.44--5.50) & ... & 1.05/ 264\\
   EM2 & 2.94 (2.91--2.98) & 1.84 (1.83--1.86) & 5.37 (5.34--5.40) & ... & 1.11/ 260\\
   EP0 & 2.74 (2.72--2.77) & 1.76 (1.75--1.77) & 5.07 (5.06--5.09) & ... & 1.13/ 623\\
   \hline
   \multicolumn{6}{c}{Joint fitting} \\
   \hline
   Soft & 2.99 (2.98--3.00) & 1.84 (1.84--1.85) & 5.69 & ... & 1.07/6217\\
   Hard & 2.99 & 2.05 (2.01--2.09) & ... & 4.87 & 0.96/  66\\
   \hline
  \end{tabular}
  \tablefoot{
  \tablefoottext{a}{The parentheses indicate the 1 $\sigma$ statistical uncertainty of
  the best-fit value. The values without the uncertainty indicate that the values
  were fixed in the fitting.}
  \tablefoottext{b}{H-equivalent column density of the interstellar extinction.}
  \tablefoottext{c}{The power-law index.}
  \tablefoottext{d}{Flux in the 2.0--8.0~keV range for the soft-band instruments.}
  \tablefoottext{e}{Flux in the 15--50~keV range for the hard-band instruments.}
  \tablefoottext{f}{The goodness of fit with the reduced $\chi^{2}$ (Red-$\chi^{2}$) and
  the degree of freedom (d.o.f.).}
  }
\end{table*}

\subsection{Individual Fitting and Systematic Uncertainties}\label{s4-2}
We now fit the spectra individually for each participating instruments following the
common procedure set up in \S~\ref{s4-1}. We justify the choice of our spectral model by
showing that all the fitting yields statistically acceptable results. We also discuss
systematic uncertainties arising from the common procedure. The spectra obtained from
the same instrument and the same response (CS0--6, SX0--1, and SX2--3) are fitted
jointly after confirming that individual spectra give consistent results.

\subsubsection{Chandra ACIS}\label{s4-2-1}
\begin{figure}[hbtp]
 \begin{center}
  \resizebox{\hsize}{!}{\includegraphics{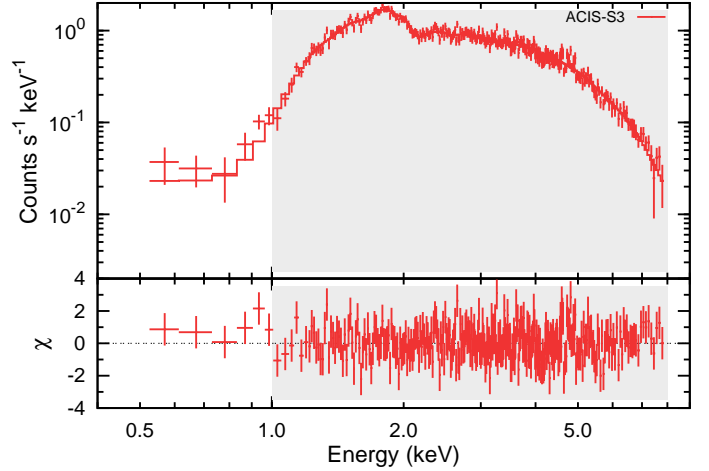}}
 \end{center}
 \caption{ACIS-S3 (CS0) spectrum and the best-fit model. The upper panel shows the data
 (crosses) and the best-fit model (solid line), while the lower panel shows the
 residuals to the fit. The range used for the fitting is shown with grey shading.}
\label{f5}
\end{figure}

The Chandra ACIS-S3 data (CS0--CS6) were taken at different epochs. G21.5--0.9 was
placed at different detector positions displaced 1\farcm17--1\farcm32 from the chip
center. The background spectra were accumulated from different regions
(figure~\ref{f2}). Statistically-acceptable best-fit models were obtained, which are
shown in figure~\ref{f5} (the CX0 fitting as a representative) and table~\ref{t2}.

\begin{figure}[hbtp]
 \begin{center}
  \resizebox{\hsize}{!}{\includegraphics{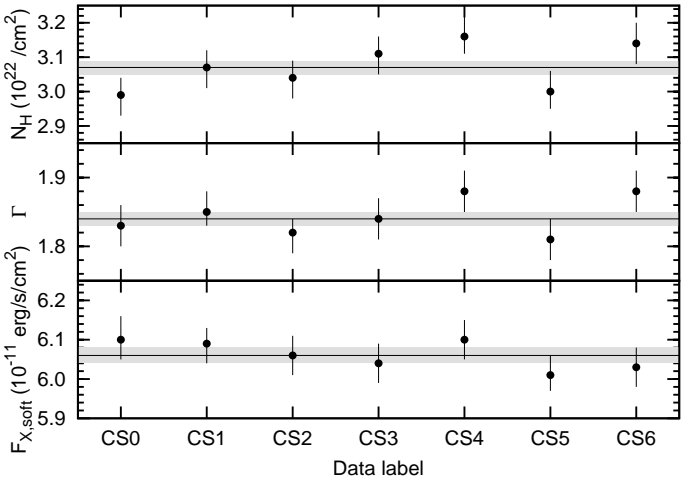}}
 \end{center}
 \caption{Comparison of best-fit values and their 1~$\sigma$ statistical uncertainty for
 the spectral parameters by the full-frame data of ACIS-S3. The solid line and the grey
 shaded regions respectively indicate the best-fit and 1~$\sigma$ statistical
 uncertainty range of joint fitting.}
 \label{f6}
\end{figure}

In order to evaluate the systematic uncertainty caused by the inhomogeneities in the
data sets, we compared the best-fit parameters among the different observations
(figure~\ref{f6}). For each parameter of $N_{\rm{H}}$, $\Gamma$, and $F_{\rm{X,soft}}$,
the seven values were tested against the null hypothesis that they deviate from a
constant value. For all parameters, the hypothesis was rejected, implying that there is
no statistical reason to claim for any difference among these data sets. We thus
combined all the spectra and conducted a joint fitting to obtain a
statistically-acceptable best-fit model. The result is appended in table~\ref{t2} as
CS0--6.

\begin{figure}[hbtp]
 \begin{center}
  \resizebox{\hsize}{!}{\includegraphics{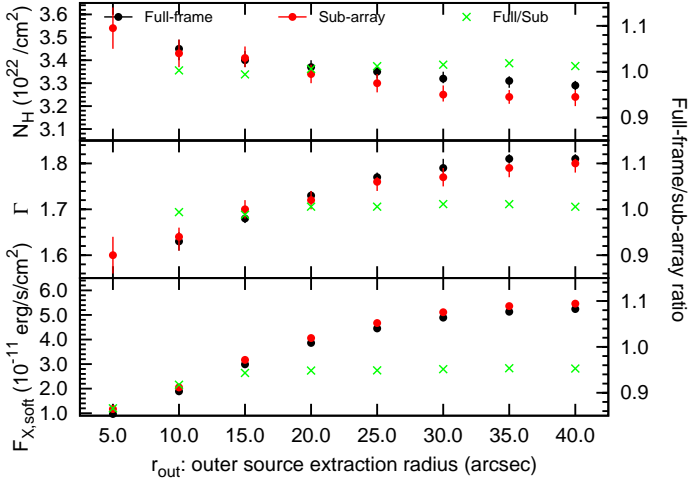}}
 \end{center}
 \caption{Comparison of best-fit values and their 1~$\sigma$ statistical uncertainty for
 the spectral parameters by joint fitting of the full-frame (black) and the sub-array
 (red) data at different outer source extraction radii. The inner radius was kept 0. The
 ratio of the two (green) are also shown for the right horizontal axis.}
 \label{f7}
\end{figure}

Due to the superb imaging capability of ACIS, the full-frame data suffer slight
pile-up at the peak of the surface brightness distribution of the object. We consider
that this effect is negligible for our cross-calibration comparison based on the
following assessment. We used the sub-array data and compared the results of the
full-frame and sub-array fitting at different outer source extraction radii
($r_\mathrm{out}$) within the field of view of the sub-array data (figure~\ref{f7}). A
shorter frame time for the sub-array data reduces the frequency of pile-up. In
figure~\ref{f7}, we see two trends as $r_\mathrm{out}$ increases. The first is the
monotonic changes of the spectral shape ($N_{\rm{H}}$ and $\Gamma$) both for the
full-frame and sub-array data, which stems from the spatial changes of the spectrum
intrinsic to G21.5--0.9. The second is the decreasing differences between the full-frame
and sub-array results, which are due to the alleviation of the pile-up effect. Beyond
20\arcsec, the results of the two data sets are stable with a full-frame to sub-array
ratio of 1 except for a 4\% difference in $F_{\rm{X,soft}}$. We consider that the
$F_{\rm{X,soft}}$ difference is a separate issue with an unidentified origin. In fact,
we found a similar discrepancy in a comparison of an annular region, in which the inner
30\arcsec\ circle including the bright core was removed to avoid pile-up. The
$F_{\rm{X,soft}}$ difference between the full-frame and sub-array data was again found at
a $\sim$3.5\% level, which exceeds the statistical uncertainty of $\sim$2.0\%. We thus
conclude that the major cause of the $F_{\rm{X,soft}}$ discrepancy is not pile-up.

\subsubsection{INTEGRAL IBIS-ISGRI}\label{s4-2-2}
\begin{figure}[hbtp]
 \begin{center}
  \resizebox{\hsize}{!}{\includegraphics{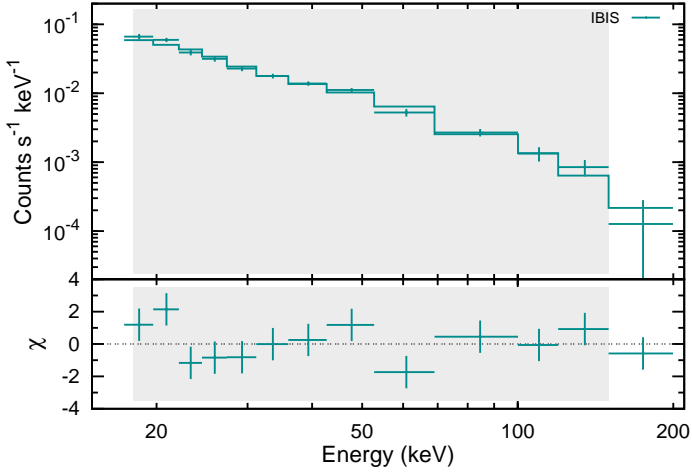}}
 \end{center}
 \caption{ISGRI spectrum and the best-fit model. The symbols follow figure~\ref{f5}.}
\label{f8}
\end{figure}

The IBIS-ISGRI detected emission from the object at a significance level of
$\sim$35\,$\sigma$. The spectrum consists of 12 channels spaced almost equally in a
logarithmic scale (figure~\ref{f8}). The interstellar extinction cannot be constrained
in the IBIS-ISGRI data alone, so we fixed the value to 2.99$\times$10$^{22}$~cm$^{-2}$,
which was obtained in the joint fitting of the soft-band instruments
(\S~\ref{s4-3-1}). A statistically-acceptable best-fit model was obtained, which is
shown in figure~\ref{f8} and table~\ref{t2}.

In order to compare with the previous work by \citet{derosa09} for the same object, we
fitted the spectrum with the same model (a single power-law) used in their work. The
derived index and flux in the 20--100~keV range is 2.0 (1.9--2.1) and 5.3 (5.1--5.5)
$\times$10$^{-11}$~erg~s$^{-1}$~cm$^{-2}$, whereas those by \citet{derosa09} are 2.2
$\pm$ 0.1 and 5.2 $\times$10$^{-11}$~erg~s$^{-1}$~cm$^{-2}$. They are consistent with
each other within the statistical uncertainty.

\subsubsection{RXTE PCA}\label{s4-2-3}
\begin{figure}[hbtp]
 \begin{center}
  \resizebox{\hsize}{!}{\includegraphics{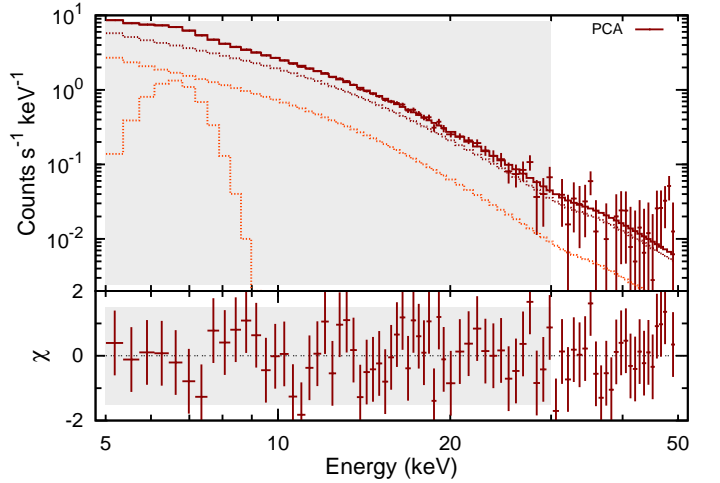}}
 \end{center}
 \caption{PCA spectrum and the best-fit model. In the upper panel, the best-fit model
 (thick brown solid line) is decomposed into the G21.5--0.9 (brown dotted line) and the
 unrelated (orange dotted line) emission components. The other symbols follow
 figure~\ref{f5}.}
 \label{f9}
\end{figure}

The non-imaging nature and the large field of view of the PCA lead to a considerable
contamination by unrelated emission ubiquitous in the direction of G21.5-0.9. Two major
sources of contamination are: (1) the Galactic ridge X-ray emission, which is evident
from an excess emission at 6--7~keV (figure~\ref{f9}) and (2) the CXB emission.

In order to subtract the contribution of the unrelated emission, we used the data
(ObsID $=$ 20266-01-35-06S) obtained in a control field devoid of bright sources at the
Galactic coordinate of ($l$, $b$) $=$ (21.999, $-$0.005), which is 1\degr.01 away from
G21.5-0.9 at (21.501, $-$0.885). We fitted the two spectra at the control field and at
G21.5--0.9 jointly. For the former, a model was applied to represent the unrelated
emission, which is a power-law continuum plus a Gaussian line for the blended Fe lines
at 6--7~keV attenuated by a fixed interstellar extinction of
$N_{\rm{H}}=$3$\times$10$^{22}$~cm$^{-2}$. For the latter, the model representing the
G21.5--0.9 spectrum was added. The interstellar extinction to G21.5--0.9 was fixed to
2.99$\times$10$^{22}$~cm$^{-2}$ similarly to the IBIS-ISGRI fitting. All parameters in
the model for the unrelated emission were common between the two spectra, except for
the flux rescaling factor for the G21.5--0.9 spectrum with respect to the control field
spectrum.

As a result, statistically-acceptable best-fit models were obtained
(figure~\ref{f9}). The best-fit parameters for the G21.5--0.9 model are summarized in
table~\ref{t2}. Those for the unrelated emission model are; $\Gamma=$2.39 (2.25--2.51),
$F_{\rm{X,hard}}=$1.50 (1.24--1.83) $\times$ 10$^{-11}$~erg~s$^{-1}$~cm$^{-2}$,
$E_{\rm{gau}}=$6.56 (6.53--6.59) keV, $\sigma_{\rm{gau}}=$0.40 (0.34--0.45) keV, and
$N_{\rm{gau}}=$3.85 (3.57--4.18) $\times$10$^{-4}$~s$^{-1}$~cm$^{-2}$, where
$E_{\rm{gau}}$, $\sigma_{\rm{gau}}$, and $N_{\rm{gau}}$ are the center energy, width,
and flux of the Gaussian model. In the 15--50~keV range, it is estimated that the
contribution of the unrelated emission accounts for 21\% of the total emission in the
G21.5-0.9 spectrum. Note that the uncertainties in the control-field flux was not
propagated to derive the numbers in table~\ref{t2}.

\subsubsection{Suzaku XIS and HXD-PIN}\label{s4-2-4}
\begin{figure}[hbtp]
 \begin{center}
  \resizebox{\hsize}{!}{\includegraphics{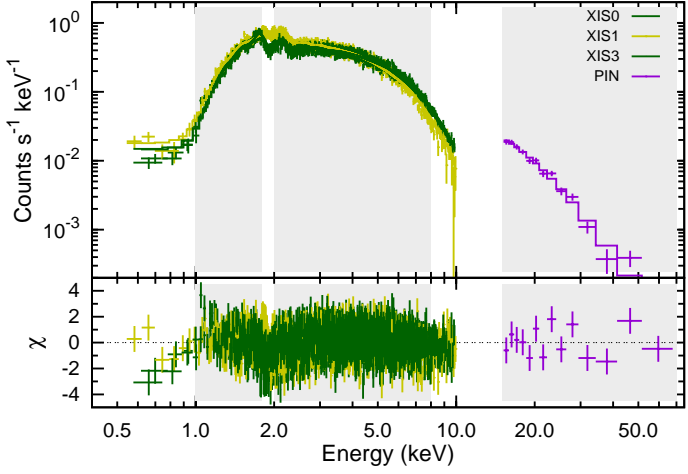}}
 \end{center}
 \caption{XIS and PIN spectra and the best-fit model. The symbols follow figure~\ref{f5}.}
\label{f10}
\end{figure}

The background events were extracted from an annulus of inner and outer radii of
5\farcm0 and 7\farcm0, which is concentric to the source extraction region. The
calibration of the XIS is uncertain around the \ion{Si}K edge. We excluded the events in
the 1.8--2.0 keV range in the fitting. We performed independent fitting for each of the
three XIS and a PIN spectra. A statistically-acceptable best-fit models were obtained,
which is shown in figure~\ref{f10} and table~\ref{t2}.

\begin{figure}[hbtp]
 \begin{center}
  \resizebox{\hsize}{!}{\includegraphics{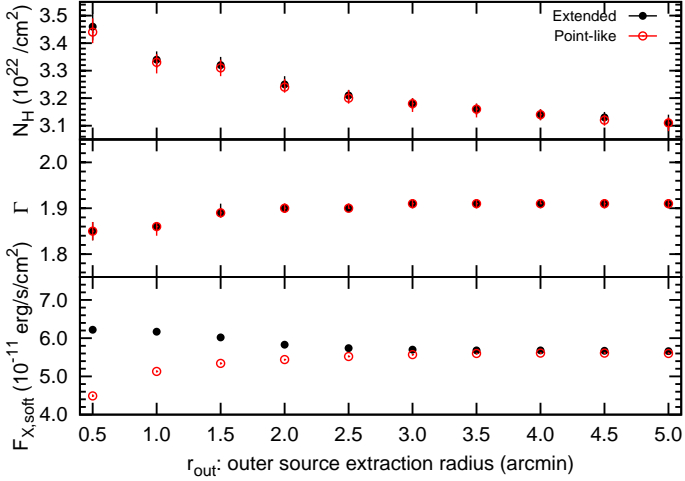}}
 \end{center}
 \caption{Best-fit values and their 1~$\sigma$ statistical uncertainty for the spectral
 parameters with different outer source extraction radii. Black symbols indicate that
 the ARF was calculated assuming the source brightness distribution of the ACIS-S3
 image. Red symbols indicate that a point-like source was assumed. Here, we use the
 result of joint fitting of the three XIS devices upon confirmation that the result of
 individual fitting gives the same trend.}
\label{f11}
\end{figure}

Unlike the other soft-band instruments, the ARF generator of the XIS compensates for
lost photos outside of the source extraction region. This is a source of systematic
uncertainty. In order to evaluate this, we fitted the spectra constructed from different
outer source extraction radii ($r_{\rm{out}}$). The inner radius was fixed to
0\arcmin. The ARFs were generated by assuming a point-like source or a surface
brightness distribution of figure~\ref{f1}. Figure~\ref{f11} summarizes the result, in
which we see the best-fit spectral parameters converge as the increasing $r_{\rm{out}}$
for both assumptions. For $r_{\rm{out}}=$5\arcmin, which we use to compare with the
other instruments, we can safely assume that the uncertainty due to the compensation of
lost photons is negligible.

\subsubsection{Swift XRT}\label{s4-2-5}
\begin{figure}[hbtp]
 \begin{center}
  \resizebox{\hsize}{!}{\includegraphics{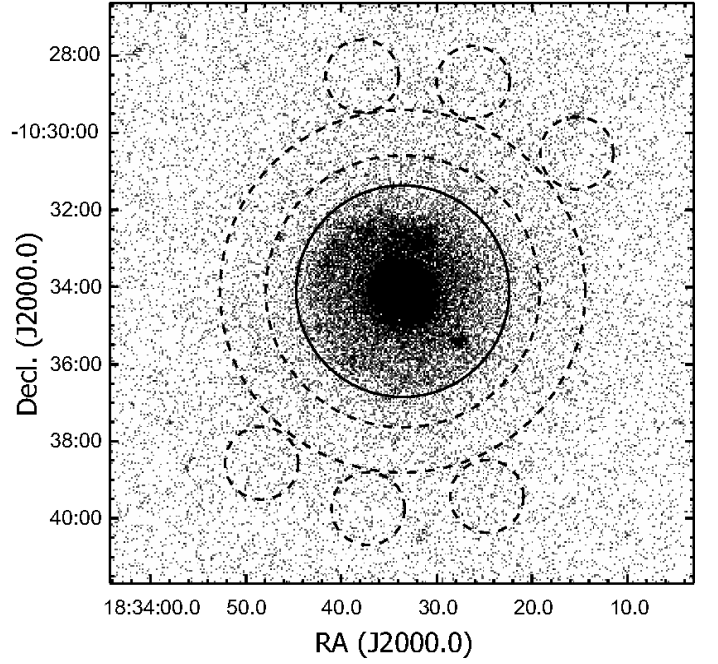}}
 \end{center}
 \caption{XRT image of G21.5--0.9. The source (solid) and background (dashed) regions
 used for the spectral analysis are shown.}
\label{f12}
\end{figure}

Figure~\ref{f12} shows the summed XRT image from which the radial profile
(figure~\ref{f4}) was calculated. Overplotted on the image are the source (solid) and
background (dashed) extraction regions. The background was chosen from an annuls and
some additional circles to increase the area from the region with a maximum and uniform
exposure.

For each of the four epochs of data (table~\ref{t1}), we performed individual
fitting. We further grouped all the data sets before and after the $V_{\rm{ss}}$ change
for joint fitting. Statistically-acceptable best-fit models were obtained, which are
shown in figure~\ref{f5} (the SX0 fitting as a representative) and table~\ref{t2}.

\begin{figure}[hbtp]
 \begin{center}
  \resizebox{\hsize}{!}{\includegraphics{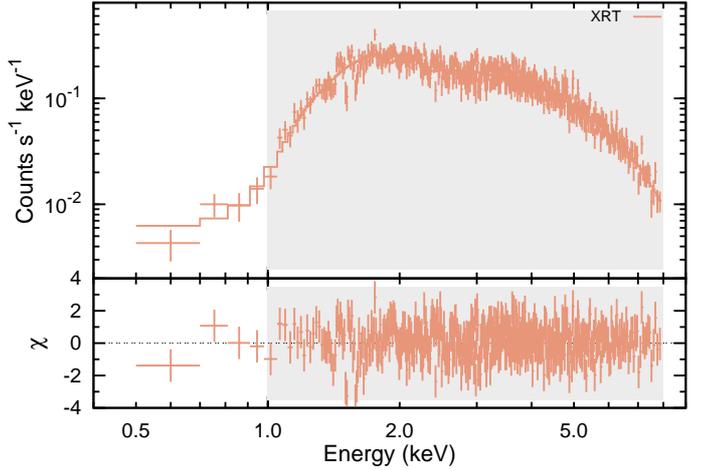}}
 \end{center}
 \caption{XRT (SX0) spectrum and the best-fit model. The symbols follow
 figure~\ref{f5}.}
\label{f13}
\end{figure}

\begin{figure}[hbtp]
 \begin{center}
  \resizebox{\hsize}{!}{\includegraphics{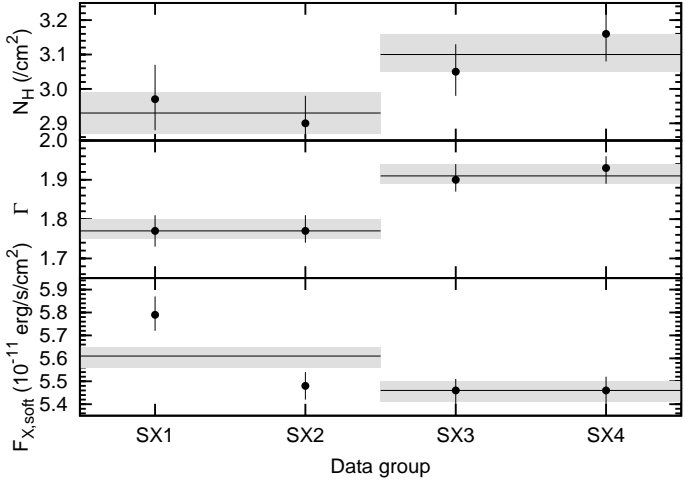}}
 \end{center}
 \caption{Comparison of best-fit values and their 1~$\sigma$ statistical uncertainty for
 the spectral parameters by XRT. The solid line and the grey shaded regions respectively
 indicate the best-fit and 1~$\sigma$ statistical uncertainty range of joint fitting.}
\label{f14}
\end{figure}

There is a clear difference in the spectral parameters obtained from data taken before
and after the $V_{\rm{ss}}$ change (figure~\ref{f14}), with the latter showing a
slightly steeper but more absorbed spectrum than the former. Within each group,
$N_{\rm{H}}$ and $\Gamma$ are consistent with each other. The discrepancy in the
parameters is caused by the slight QE change incurred after the $V_{\rm{ss}}$ was
altered. The calibration of this effect is ongoing. We hereafter use the result obtained
for the data before the change (SX0 and SX1). We see a notable difference in
$F_{\rm{X,soft}}$ between SX0 and SX1. We could not identify the cause for the
difference and ignore it as a statistical scatter.

\subsubsection{XMM-Newton EPIC}\label{s4-2-6}
In order to ensure that the fluxes measured by the three EPIC cameras correspond to the
same region in the sky, as well as to provide a correction factor for the lost sky area
due to bad pixels/columns and chip gaps, we applied a common detector mask during the
accumulation of the spectral products (``masked spectra'' hereafter). This mask was
generated by multiplying the detector masks produced by the SAS task \texttt{emask} on
the individual EPIC cameras\footnote{SAS users should be aware that this procedure does
not work with SAS versions 10 and earlier. Due to a formal error in the header of the
mask file, the mask description is not correctly propagated in the spectral file header,
and \texttt{arfgen} cannot properly apply the bad area correction on ''masked''
spectra. We modified the header of the mask file manually in order to cope with this
problem. A solution to this problem is expected to be available in SAS versions later
then 10.0.}. For the correction factor to compensate for the masked area, we extracted a
MOS1 spectrum from the full 165\arcsec\ extraction circle (''unmasked spectrum''). We
chose the MOS1 detector here, because it was only marginally affected by CCD artefacts
in our data. We calculated the flux ratio between the unmasked and the masked spectra,
which was 1.091 in the 2--8~keV energy range. Finally, we multiplied the ratio to the
flux obtained by the ``masked'' spectra to derive the flux in the full 165\arcsec\
circle.

The background spectra for the two MOS cameras were extracted from an annulus of the
inner and outer radii of 3\farcm3 and 5\farcm0, which is concentric to the source
extraction region. The background region is also contained in the central CCD chip with
the source region (figure~\ref{f3}).

Unlike the MOS cameras, the background extraction for the pn camera is not
straight-forward. The source spreads across multiple CCDs (figure~\ref{f3}) and the
background level is known to have a measurable position dependence, in particular,
along the readout direction. We therefore extracted a spectrum from blank sky data
\citep{carter07} using the same extraction region with the source and used it as the
background spectrum. Out-of-time events were removed in the pn data.

\begin{figure}[hbtp]
 \begin{center}
  \resizebox{\hsize}{!}{\includegraphics{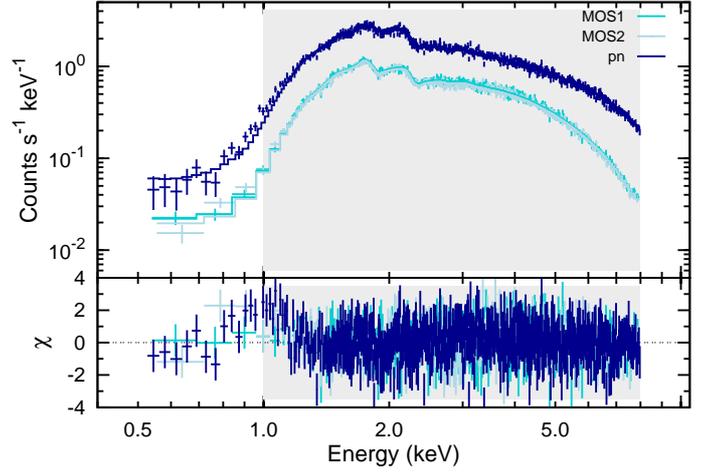}}
 \end{center}
 \caption{EPIC spectra and the best-fit model. The symbols follow figure~\ref{f5}.}
\label{f15}
\end{figure}

We fitted individual spectra of the three EPIC devices. A statistically-acceptable
best-fit model was obtained, which is shown in figure~\ref{f15} and table~\ref{t2}.

\begin{figure}[hbtp]
 \begin{center}
  \resizebox{\hsize}{!}{\includegraphics{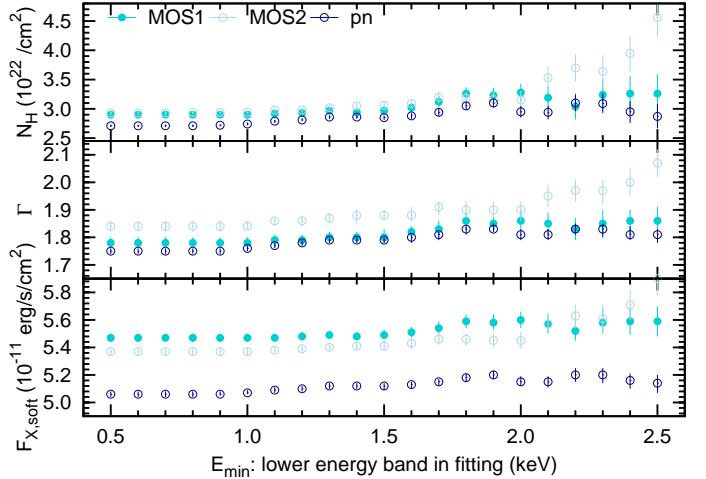}}
 \end{center}
 \caption{Best-fit values and their 1~$\sigma$ statistical uncertainty for the spectral
 parameters with different lower energy band used in the fitting.}
\label{f16}
\end{figure}

In the pn result, a deviation is seen below 1.5~keV, which is not observed in any other
instruments at a comparable level. The origin of this feature is currently
unknown. Possible explanations are residual uncertainties in the background subtraction
(the pn background has a complex spatial structure whose modelling for extended sources
close to the bore-sight position is non-trivial) and residual inaccuracies in the
calibration of the redistribution. Notwithstanding the origin of this feature, we
assessed the effect of the deviation to the best-fit parameter values by altering the
lower energy ($E_{\rm{min}}$) in the fitting (figure~\ref{f16}). The upper energy
($E_{\rm{max}}$) was fixed to 8.0~keV. Toward lower energies in $E_{\rm{min}}$, the
best-fit $N_{{\rm{H}}}$ and $\Gamma$ values change monotonically. The global trend,
however, is common among MOS1, MOS2, and pn, implying that the $<$1.5~keV deviation only
seen in the pn does not affect the pn fitting results. This is conceivable given the
fact that the pn spectrum has dominant counts in 1.5--$E_{\rm{max}}$ keV band. We thus
ignore the effect of the $<$1.5~keV deviation hereafter.

\subsection{Joint Fitting of All Data}\label{s4-3}
We assembled all the spectra and responses and attempted a joint fitting separately for
the soft-band and hard-band instruments. The procedures follow the individual fitting in
\S~\ref{s4-2}. The seven ACIS-S3 spectra were treated as one. The two XRT data (SX2 and
SX3) were discarded and the remaining two (SX0 and SX1) were used as one.

\subsubsection{Soft-band Instruments}\label{s4-3-1}
In the soft-band joint fitting, we first fitted the spectra with all the spectral
parameters ($N_{\rm{H}}$, $\Gamma$, $F_{\rm{X,soft}}$) tied among all the
instruments. The best-fit model was statistically rejected due mainly to the
inconsistent normalizations among the instruments. The statistical uncertainties were not
derived as the goodness of fit was very low.

\begin{figure}[hbtp]
 \begin{center}
  \resizebox{\hsize}{!}{\includegraphics{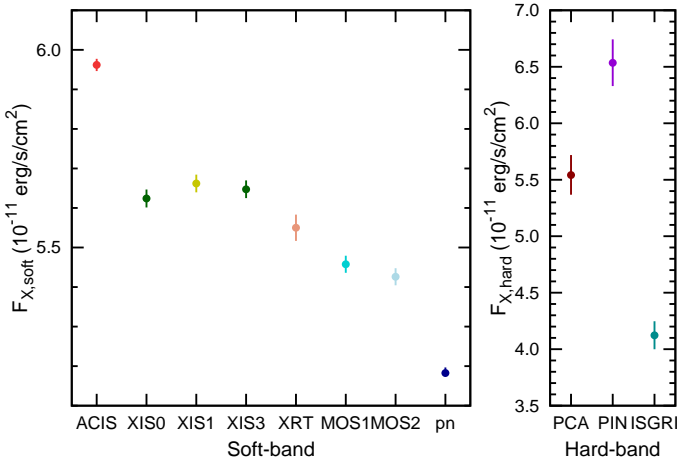}}
 \end{center}
 \caption{Comparison of the best-fit $F_{\rm{X,soft}}$ or $F_{\rm{X,hard}}$ values and
 their 1 $\sigma$ uncertainty in the joint fitting with the other parameters ($\Gamma$
 and $N_{H}$) tied among all the soft-band or hard instruments.}
\label{f17}
\end{figure}

We thus introduced an additional normalization rescaling parameter for all the
instruments with respect to a fixed $F_{\rm{X,soft}}$ value of
5.69$\times$10$^{-11}$~erg~s$^{-1}$~cm$^{-2}$, which is the best-fit value in the joint
fitting above. Note that this value is not the most likely value for the absolute
flux. The derived value can be easily biased by the instruments yielding a larger number
of counts than the others regardless of whether they are well-calibrated or
not. Nevertheless, it is a value that best represents our datasets. We obtained a
result, which was still statistically unacceptable but was much improved in the goodness
of fit (table~\ref{t2}). The comparison of the $F_{\rm{X,soft}}$ is shown in
figure~\ref{f17}.

\subsubsection{Hard-band Instruments}\label{s4-3-2}
We took the same approach in the joint fitting of the hard-band instruments except that
the $N_{H}$ value was fixed to 2.99$\times$10$^{22}$~cm$^{-2}$, which is the result of
the soft-band joint fitting. First, we fitted all the spectral parameters ($\Gamma$,
$F_{\rm{X,hard}}$) tied among all the instruments, which was in vain for the same reason
with the soft-band joint fitting. We thus introduced the normalization rescaling
parameter for each spectrum and obtained a statistically-acceptable best-fit model
(table~\ref{t2}). The comparison of the $F_{\rm{X,hard}}$ is shown in figure~\ref{f17}.

\section{DISCUSSION}\label{s5}
\subsection{Comparison of Fitting Results}
\begin{figure}[hbtp]
 \begin{center}
  \resizebox{\hsize}{!}{\includegraphics{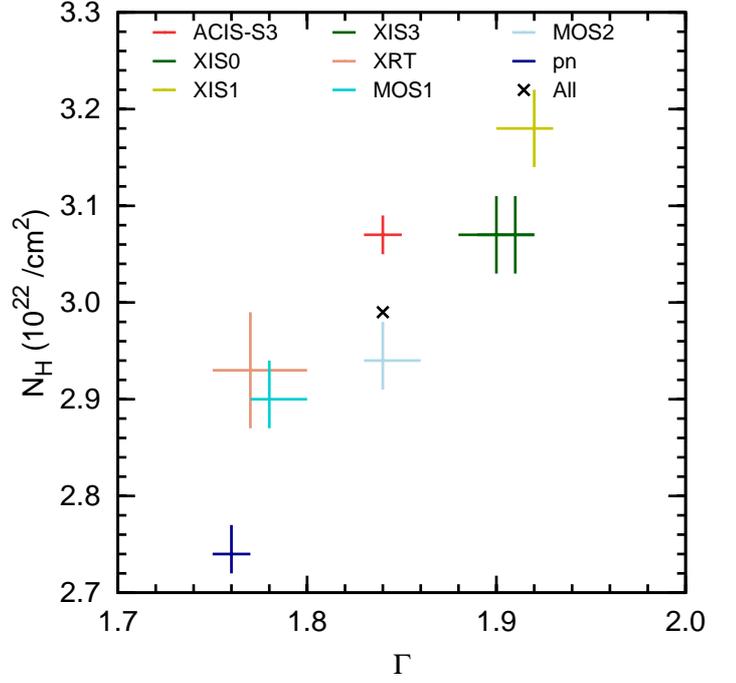}}
 \end{center}
 \caption{Scatter plot of the best-fit $\Gamma$ versus $N_{\rm{H}}$ values and their
 1\,$\sigma$ statistical uncertainty of all soft-band instruments.}
\label{f18}
\end{figure}

\begin{figure}[hbtp]
 \begin{center}
  \resizebox{\hsize}{!}{\includegraphics{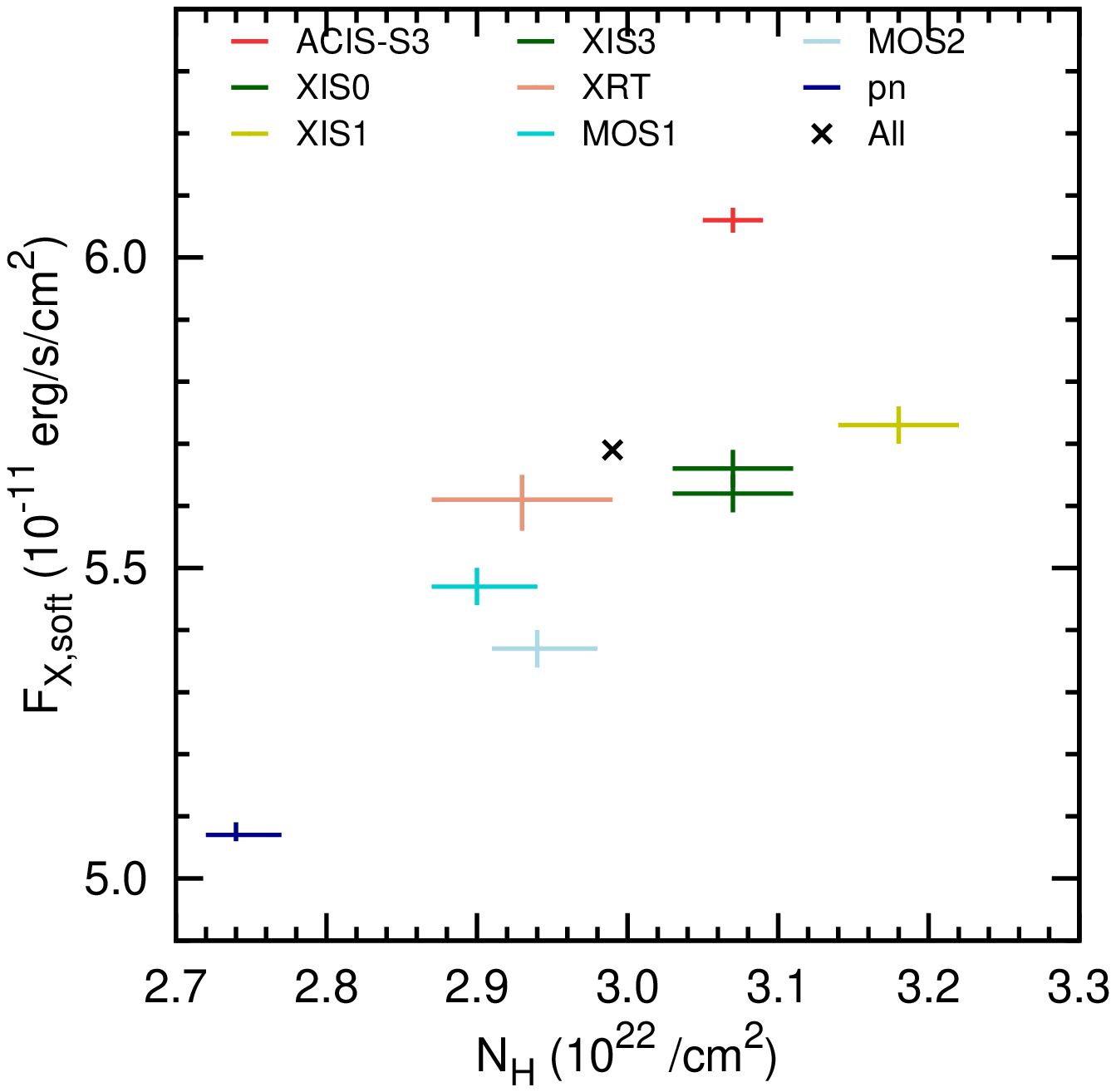}}
 \end{center}
 \caption{Scatter plot of the best-fit $N_{\rm{H}}$ versus $F_{\rm{X,soft}}$ values and
 their 1\,$\sigma$ statistical uncertainty of all soft-band instruments.}
\label{f19}
\end{figure}

\begin{figure}[hbtp]
 \begin{center}
  \resizebox{\hsize}{!}{\includegraphics{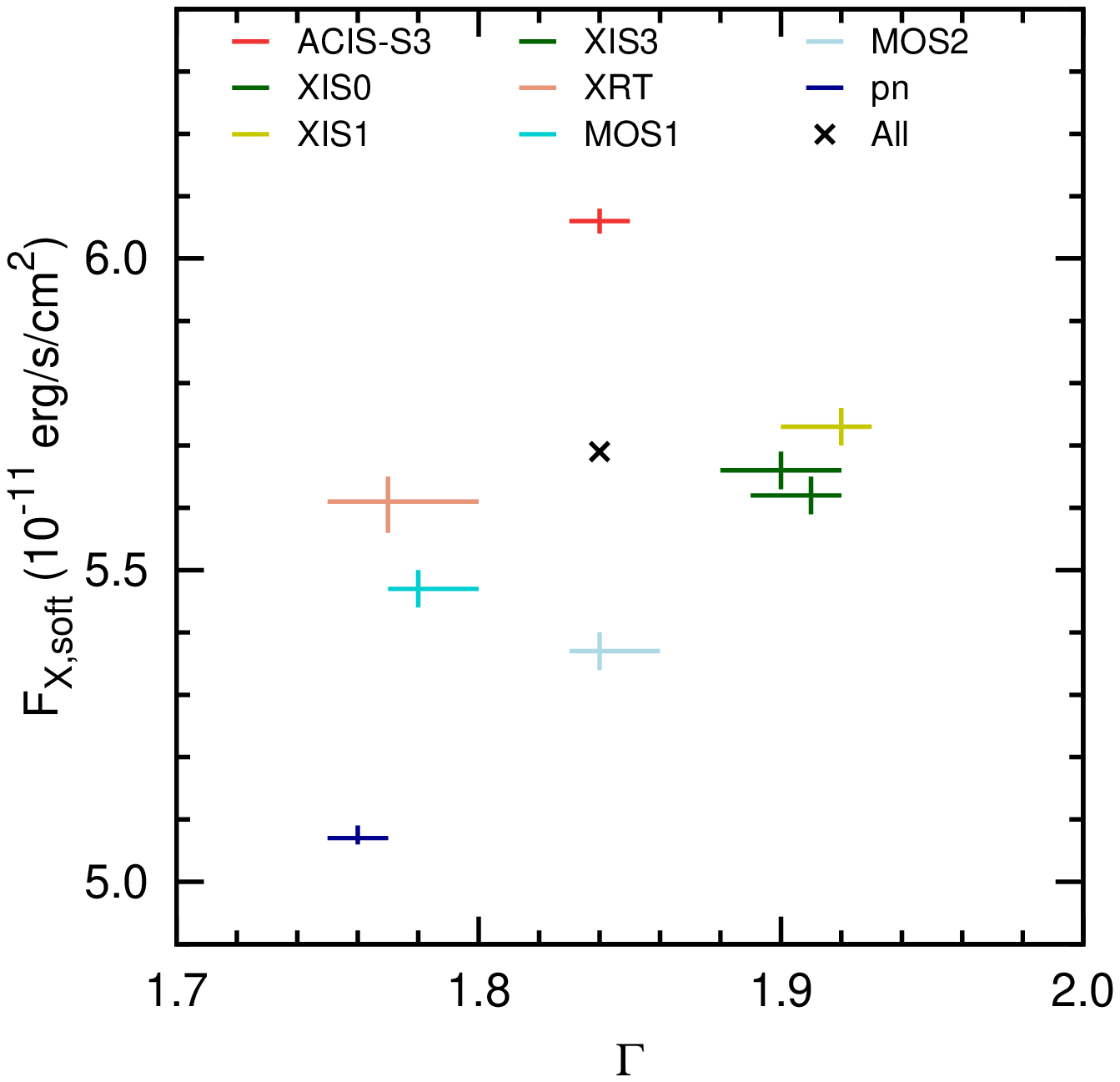}}
 \end{center}
 \caption{Scatter plot of the best-fit $\Gamma$ versus $F_{\rm{X,soft}}$ values and
 their 1\,$\sigma$ statistical uncertainty of all soft-band instruments.}
\label{f20}
\end{figure}

\begin{figure}[hbtp]
 \begin{center}
  \resizebox{\hsize}{!}{\includegraphics{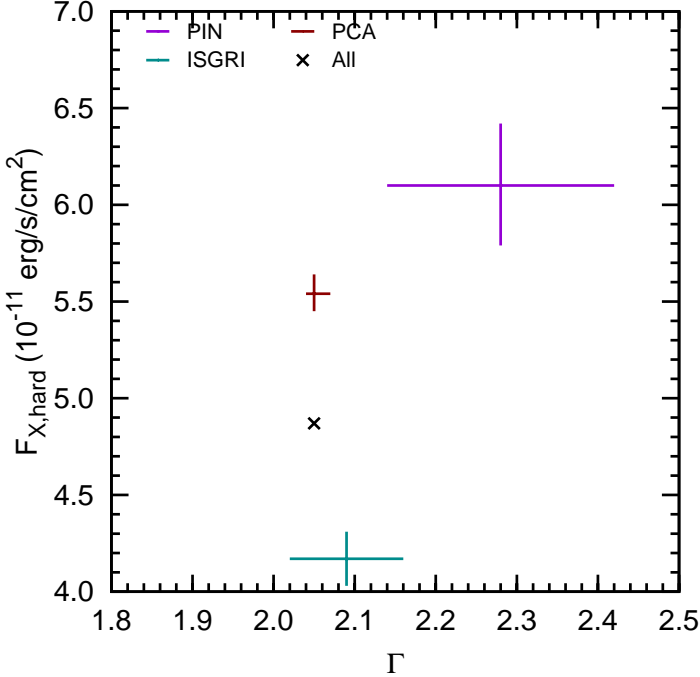}}
 \end{center}
 \caption{Scatter plot of the best-fit $\Gamma$ versus $F_{\rm{X,hard}}$values and their
 1\,$\sigma$ statistical uncertainty of all hard-band instruments.}
\label{f21}
\end{figure}

We now compare all the results. The values in table~\ref{t2} are plotted in four scatter
plots (figure~\ref{f18}, \ref{f19}, \ref{f20}, and \ref{f21}) for every combination of
two spectral parameters.

\subsubsection{Identifying Systematic Differences}
For identifying significant systematic uncertainties, we assume that the instrument A and
B respectively have the best-fit value and 1 $\sigma$ statistical uncertainty for a
certain spectral parameter $x$ as $x^{\mathrm{(A)}} \pm \Delta x^{\mathrm{(A)}}$ and
$x^{\mathrm{(B)}} \pm \Delta x^{\mathrm{(B)}}$. The statistical uncertainty is nearly
symmetric in the upper and lower directions in all parameters (table~\ref{t2}), so we
took the mean of the two for $\Delta x^{\mathrm{(A)}}$ and $\Delta x^{\mathrm{(B)}}$
here. If the modulus of the logarithm of the ratio exceeds three times the convolved
statistical uncertainty as
\begin{equation}
\left|\ln{\left(\frac{x^{\mathrm{(B)}}}{x^{\mathrm{(A)}}}\right)}\right| > 3\sqrt{\left(\frac{\Delta x^{\mathrm{(A)}}}{x^{\mathrm{(A)}}}\right)^{2} + \left(\frac{\Delta x^{\mathrm{(B)}}}{x^{\mathrm{(B)}}}\right)^{2}},
\end{equation}
we consider that the two instruments A and B have a significant systematic difference in
$x$.

In the soft band, we have eight instruments (ACIS-S3, XIS0, XIS1, XIS3, XRT, MOS1, MOS2,
and pn) and three parameters ($N_{\rm{H}}$, $\Gamma$, and
$F_{\rm{X,soft}}$). Figure~\ref{f18} shows that $N_{\rm{H}}$ and $\Gamma$ are coupled,
so we only consider $\Gamma$. For each parameter, we have 28 combinations for a pair of
instruments. In the hard band, we have three instruments (PIN, PCA, and ISGRI) and two
parameters ($\Gamma$ and $F_{\rm{X,hard}}$). For each parameter, we have 3
combinations. A total of 62 combinations were examined. The identified systematic
differences are shown in bold faces in tables~\ref{t3}, \ref{t4}, \ref{t5}, and
\ref{t6}.

\subsubsection{Soft-band Comparison}
In the flux comparison, 20 out of 28 combinations show a significant difference
(table~\ref{t3}). The ACIS-S3 flux is larger than all the others. The pn flux is smaller
than all the others. The three XIS instruments (XIS0, XIS1, and XIS3) are consistent
with each other, while the two MOS cameras are consistent with each other.  XRT is
consistent with the three XIS devices and MOS1. The largest difference, between ACIS-S3
and pn, is 20\%.

In the index comparison, 16 out of 28 combinations show a significant difference
(table~\ref{t4}). The three XIS instruments and the two MOS cameras are consistent with
each other also in $\Gamma$. The largest difference, between XIS1 and pn, is 9\%.

\begin{table*}[tpbh]
 \caption{Relative flux among soft-band instruments.\tablefootmark{a}}\label{t3}
 \centering
 \begin{tabular}{|cc|cccccccc|}
  \hline 
  &       & \multicolumn{8}{|c|}{A} \\
  \cline{3-10}
  &       & ACIS-S3 & XIS0 & XIS1 & XIS3 & XRT & MOS1 & MOS2 & pn \\
  \hline 
  & \multicolumn{1}{|c|}{ACIS-S3} 
  & 0.00
  & \textbf{7.54 $\pm$ 0.63}
  & \textbf{5.60 $\pm$ 0.62}
  & \textbf{6.83 $\pm$ 0.62}
  & \textbf{7.72 $\pm$ 0.87}
  & \textbf{10.24 $\pm$ 0.64}
  & \textbf{12.09 $\pm$ 0.65}
  & \textbf{17.84 $\pm$ 0.44}
  \\
  & \multicolumn{1}{|c|}{XIS0}
  & \textbf{--7.54 $\pm$ 0.63}
  & 0.00
  & --1.94 $\pm$ 0.75
  & --0.71 $\pm$ 0.75
  & 0.18 $\pm$ 0.96
  & \textbf{2.71 $\pm$ 0.77}
  & \textbf{4.55 $\pm$ 0.77}
  & \textbf{10.30 $\pm$ 0.61}
  \\
  & \multicolumn{1}{|c|}{XIS1}
  & \textbf{--5.60 $\pm$ 0.62}
  & 1.94 $\pm$ 0.75
  & 0.00
  & 1.23 $\pm$ 0.74
  & 2.12 $\pm$ 0.96
  & \textbf{4.64 $\pm$ 0.76}
  & \textbf{6.49 $\pm$ 0.77}
  & \textbf{12.24 $\pm$ 0.60}
  \\
  & \multicolumn{1}{|c|}{XIS3}
  & \textbf{--6.83 $\pm$ 0.62}
  & 0.71 $\pm$ 0.75
  & --1.23 $\pm$ 0.74
  & 0.00
  & 0.89 $\pm$ 0.96
  & \textbf{3.41 $\pm$ 0.76}
  & \textbf{5.26 $\pm$ 0.77}
  & \textbf{11.01 $\pm$ 0.61}
  \\
 B& \multicolumn{1}{|c|}{XRT}
  & \textbf{--7.72 $\pm$ 0.87}
  & --0.18 $\pm$ 0.96
  & --2.12 $\pm$ 0.96
  & --0.89 $\pm$ 0.96
  & 0.00
  & 2.53 $\pm$ 0.97
  & \textbf{4.37 $\pm$ 0.98}
  & \textbf{10.12 $\pm$ 0.85}
  \\
  & \multicolumn{1}{|c|}{MOS1}
  & \textbf{--10.24 $\pm$ 0.64}
  & \textbf{--2.71 $\pm$ 0.77}
  & \textbf{--4.64 $\pm$ 0.76}
  & \textbf{--3.41 $\pm$ 0.76}
  & --2.53 $\pm$ 0.97
  & 0.00
  & 1.85 $\pm$ 0.78
  & \textbf{7.59 $\pm$ 0.62}
  \\
  & \multicolumn{1}{|c|}{MOS2}
  & \textbf{--12.09 $\pm$ 0.65}
  & \textbf{--4.55 $\pm$ 0.77}
  & \textbf{--6.49 $\pm$ 0.77}
  & \textbf{--5.26 $\pm$ 0.77}
  & \textbf{--4.37 $\pm$ 0.98}
  & --1.85 $\pm$ 0.78
  & 0.00
  & \textbf{5.75 $\pm$ 0.63}
  \\
  & \multicolumn{1}{|c|}{pn}
  & \textbf{--17.84 $\pm$ 0.44}
  & \textbf{--10.30 $\pm$ 0.61}
  & \textbf{--12.24 $\pm$ 0.60}
  & \textbf{--11.01 $\pm$ 0.61}
  & \textbf{--10.12 $\pm$ 0.85}
  & \textbf{--7.59 $\pm$ 0.62}
  & \textbf{--5.75 $\pm$ 0.63}
  & 0.00
  \\
  \hline 
 \end{tabular}
 \tablefoot{
 \tablefoottext{a}{The logarithmic flux ratio $100 \times
 \ln{\left(F^{\mathrm{(B)}}_{\mathrm{X}}/F^{\mathrm{(A)}}_{\mathrm{X}}\right)}$ between
 the soft-band instruments A and B. The flux is measured in the 2.0--8.0~keV band. The
 range indicates the convolved statistical uncertainty derived as $100 \times
 \sqrt{\left(\frac{\Delta
 F^{\mathrm{(A)}}_{\mathrm{X}}}{F^{\mathrm{(A)}}_{\mathrm{X}}}\right)^{2}+\left(\frac{\Delta
 F^{\mathrm{(B)}}_{\mathrm{X}}}{F^{\mathrm{(B)}}_{\mathrm{X}}}\right)^{2}}$. Both the
 ratio and its deviation are multiplied by 100 to save space. The uncertainties $\Delta
 F^{\mathrm{(A)}}_{\mathrm{X}}$ and $\Delta F^{\mathrm{(B)}}_{\mathrm{X}}$ are the mean
 of the 1 $\sigma$ statistical uncertainties in the upper and lower bound directions
 (table~\ref{t2}). The bold face indicates that the difference is larger than 3 times
 the convolved statistical uncertainty.}
 }
\end{table*}

\begin{table*}[tpbh]
 \caption{Relative index of power among soft-band instruments.\tablefootmark{a}}\label{t4}
 \centering
 \begin{tabular}{|cc|cccccccc|}
  \hline 
  &       & \multicolumn{8}{|c|}{A} \\
  \cline{3-10}
  &       & ACIS-S3 & XIS0 & XIS1 & XIS3 & XRT & MOS1 & MOS2 & pn \\
  \hline 
  & \multicolumn{1}{|c|}{ACIS-S3} 
  & 0.00
  & \textbf{--3.73 $\pm$ 0.96}
  & \textbf{--4.26 $\pm$ 0.95}
  & --3.21 $\pm$ 1.18
  & 3.88 $\pm$ 1.51
  & \textbf{3.32 $\pm$ 1.00}
  & 0.00 $\pm$ 0.98
  & \textbf{4.45 $\pm$ 0.79}
  \\
  & \multicolumn{1}{|c|}{XIS0}
  & \textbf{3.73 $\pm$ 0.96}
  & 0.00
  & --0.52 $\pm$ 1.11
  & 0.52 $\pm$ 1.31
  & \textbf{7.61 $\pm$ 1.62}
  & \textbf{7.05 $\pm$ 1.15}
  & \textbf{3.73 $\pm$ 1.13}
  & \textbf{8.18 $\pm$ 0.97}
  \\
  & \multicolumn{1}{|c|}{XIS1}
  & \textbf{4.26 $\pm$ 0.95}
  & 0.52 $\pm$ 1.11
  & 0.00
  & 1.05 $\pm$ 1.31
  & \textbf{8.13 $\pm$ 1.61}
  & \textbf{7.57 $\pm$ 1.15}
  & \textbf{4.26 $\pm$ 1.13}
  & \textbf{8.70 $\pm$ 0.97}
  \\
  & \multicolumn{1}{|c|}{XIS3}
  & 3.21 $\pm$ 1.18
  & --0.52 $\pm$ 1.31
  & --1.05 $\pm$ 1.31
  & 0.00
  & \textbf{7.09 $\pm$ 1.76}
  & \textbf{6.52 $\pm$ 1.35}
  & 3.21 $\pm$ 1.33
  & \textbf{7.65 $\pm$ 1.20}
  \\
 B& \multicolumn{1}{|c|}{XRT}
  & --3.88 $\pm$ 1.51
  & \textbf{--7.61 $\pm$ 1.62}
  & \textbf{--8.13 $\pm$ 1.61}
  & \textbf{--7.09 $\pm$ 1.76}
  & 0.00
  & --0.56 $\pm$ 1.64
  & --3.88 $\pm$ 1.63
  & 0.57 $\pm$ 1.52
  \\
  & \multicolumn{1}{|c|}{MOS1}
  & \textbf{--3.32 $\pm$ 1.00}
  & \textbf{--7.05 $\pm$ 1.15}
  & \textbf{--7.57 $\pm$ 1.15}
  & \textbf{--6.52 $\pm$ 1.35}
  & 0.56 $\pm$ 1.64
  & 0.00
  & --3.32 $\pm$ 1.17
  & 1.13 $\pm$ 1.02
  \\
  & \multicolumn{1}{|c|}{MOS2}
  & 0.00 $\pm$ 0.98
  & \textbf{--3.73 $\pm$ 1.13}
  & \textbf{--4.26 $\pm$ 1.13}
  & --3.21 $\pm$ 1.33
  & 3.88 $\pm$ 1.63
  & 3.32 $\pm$ 1.17
  & 0.00
  & \textbf{4.45 $\pm$ 0.99}
  \\
  & \multicolumn{1}{|c|}{pn}
  & \textbf{--4.45 $\pm$ 0.79}
  & \textbf{--8.18 $\pm$ 0.97}
  & \textbf{--8.70 $\pm$ 0.97}
  & \textbf{--7.65 $\pm$ 1.20}
  & --0.57 $\pm$ 1.52
  & --1.13 $\pm$ 1.02
  & \textbf{--4.45 $\pm$ 0.99}
  & 0.00
  \\
  \hline 
 \end{tabular}
 \tablefoot{
 \tablefoottext{a}{The logarithmic index of power ratio
 $100 \times \ln{\left(\Gamma^{\rm{(B)}}/\Gamma^{\rm{(A)}}\right)}$  between the soft-band
 instruments A and B. The notation follows table~\ref{t3}.}
 }
\end{table*}

\subsubsection{Hard-band Comparison}
In the flux comparison, 2 out of 3 combinations in $F_{\rm{X,hard}}$ are significantly
different (table~\ref{t5}). ISGRI is smaller than the other two. The largest difference
between ISGRI and PIN is as large as 46\%. In the index comparison, all three
instruments are consistent with each other (table~\ref{t6}).

\begin{table*}[tpbh]
 \caption{Relative flux among hard-band instruments.\tablefootmark{a}}\label{t5}
 \centering
 \begin{tabular}{|cc|ccc|}
  \hline 
  &       & \multicolumn{3}{|c|}{A} \\
  \cline{3-5}
  &       & ISGRI& PCA & PIN \\
  \hline 
  & \multicolumn{1}{|c|}{ISGRI}
  & 0.00
  & \textbf{--28.41 $\pm$ 3.77}
  & \textbf{--38.04 $\pm$ 6.16}
  \\
 B& \multicolumn{1}{|c|}{PCA}
  & \textbf{28.41 $\pm$ 3.77}
  & 0.00
  & --9.63 $\pm$ 5.44
  \\
  & \multicolumn{1}{|c|}{PIN}
  & \textbf{38.04 $\pm$ 6.16}
  & 9.63 $\pm$ 5.44
  & 0.00
  \\
  \hline 
 \end{tabular}
 \tablefoot{
 \tablefoottext{a}{The logarithmic flux ratio
 $100 \times \ln{\left(F^{\rm{(B)}}_{\rm{X}}/F^{\rm{(A)}}_{\rm{X}}\right)}$ between the hard-band
 instruments A and B. The flux is measured in the 15--50.0~keV band. The notation
 follows table~\ref{t3}.}
 }
\end{table*}

\begin{table*}[tpbh]
 \caption{Relative index of power among hard-band instruments.\tablefootmark{a}}\label{t6}
 \centering
 \begin{tabular}{|cc|ccc|}
  \hline 
  &       & \multicolumn{3}{|c|}{A} \\
  \cline{3-5}
  &       & ISGRI & PCA & PIN \\
  \hline 
  & \multicolumn{1}{|c|}{ISGRI}
  & 0.00
  & 1.93 $\pm$ 3.43
  & --8.70 $\pm$ 6.99
  \\
 B& \multicolumn{1}{|c|}{PCA}
  & --1.93 $\pm$ 3.43
  & 0.00
  & --10.63 $\pm$ 6.18
  \\
  & \multicolumn{1}{|c|}{PIN}
  & 8.70 $\pm$ 6.99
  & 10.63 $\pm$ 6.18
  & 0.00
  \\
  \hline 
 \end{tabular}
 \tablefoot{
 \tablefoottext{a}{The logarithmic index of power ratio
 $100 \times \ln{\left(\Gamma^{\rm{(B)}}/\Gamma^{\rm{(A)}}\right)}$
 between the hard-band instruments A and B. The notation follows table~\ref{t3}.}
 }
\end{table*}

\subsection{Comparison with Other Studies}
\subsubsection{Internal Cross-calibration Studies}
We compared our results with the cross-calibration results for the instruments onboard
the same satellite (i.e., XIS0, XIS1, and XIS3 for Suzaku and EPIC-MOS1, EPIC-MOS2, and
EPIC-pn for the XMM-Newton), referring to the technical notes issued by their
calibration teams. Both observatories use these instruments simultaneously, so we can
mitigate the effects caused by the changes of the instrumental performance in time in
these comparisons.

In the Suzaku XIS study \citep{ishida07}, the Crab nebula was used to compare the
power-law index and the flux in 1--10~keV at the XIS nominal position. Their results are
consistent with the present study. XIS0, 1, 3 are consistent with each other to better
than 3\% both in the flux and power-law index. The back-side illuminated device (XIS1)
is slightly larger in flux and steeper in the power-law index than the front-side
illuminated devices (XIS0 and XIS3).

In the XMM-Newton EPIC study \citep{gabriel08}, G21.5--0.9 was also used to compare the
power-law index and the flux in 2--10~keV band using the same data set. Their results
shows that the flux increases in the sequence of MOS2, MOS1, and pn, which is
inconsistent with ours. This is due to the caveat of the bad pixel/column treatment in
the older SAS versions described in \S~\ref{s4-2-6}. In fact, without considering this
caveat, we obtained the same result with \citet{gabriel08}.

In another XMM-Newton EPIC study \citep{stuhlinger08}, several extra-galactic point-like
sources are used to compare the flux at a half dozen energy bands. Point sources are
less affected by the SAS caveat than extended sources like G21.5--0.9. In the two energy
bands closest to ours (1.5--4.0 and 4.0--10.0~keV), the flux is larger in MOS1 and MO2
than pn by $\sim$5\%, which is the same trend with our result.

\subsubsection{External Cross-calibration Studies}
We also compared our results with other cross-calibration studies across multiple
missions. The number of such studies is actually small; in fact, we could only find
several studies comparing Chandra and XMM-Newton often with outdated software
versions. We refer to another IACHEC study \cite{nevalainen10}, in which 11 relaxed
clusters of galaxies are used.

The 2--7~keV band spectra were fitted with an optically-thin thermal plasma model to
derive the plasma temperature and the flux. The ACIS-I, ACIS-S, and EPIC instruments
were used. ACIS-I and ACIS-S were confirmed to be consistent with each other and were
treated as one. 

In the flux comparison, ACIS was significantly larger than the three EPIC cameras by
5--10\%. Among the three EPIC instruments, pn is lower than the other two by
$\sim$5\%. Both discrepancies are consistent with our result.

In the temperature comparison, ACIS and pn were consistent with each other. Among the
EPIC instruments, the pn and the MOS (MOS1 and MOS2 were combined) were also consistent
with each other. In order to compare to the cluster temperature result, we fitted the
G21.5--0.9 spectra in the same energy band with a power-law model attenuated by a fixed
$N_{\rm{H}}$ value of 2.99$\times$10$^{22}$~cm$^{-2}$ and substituted the softness of
the index with the temperature. We found that the consistency in the spectral softness
among the ACIS-S3 and the three EPIC cameras is also found in G21.5--0.9.

\section{CONCLUSION}\label{s6}
We used the pulsar wind nebula G21.5--0.9 to conduct a cross-calibration study of
instruments onboard the currently working X-ray astronomy missions. The archived as well
as the original data were accumulated for Chandra ACIS-S3, INTEGRAL IBIS-ISGRI, RXTE
PCA, Suzaku XIS and HXD-PIN, Swift XRT, and XMM-Newton EPIC-MOS and EPIC-pn.

We conducted a coherent spectral fitting for all the data with the methods as common as
possible. We tabulated and plotted the results to be a useful reference. We compared the
results in all combinations of two different instruments in the spectral parameters and
identified many systematic differences unattributable to the statistical uncertainties
alone. Most differences are consistent with those reported in previous studies of a
similar purpose.

Many factors can cause the systematic differences among different instruments, including
(1) inaccurate calibration of either or both of the two compared instruments, (2) the
limitations of G21.5--0.9 as a calibration source, and (3) inappropriate choice of the
source extraction aperture, the spectral models, background, energy ranges, etc. These
are entangled with each other, and our aim is to discriminate the causes belonging to
(1). We addressed some of the concerns belonging to (3) in \S~\ref{s4-2} and showed that
their effects are minor. The limitations of G21.5--0.9 as a calibration source
(\S~\ref{s2}) are unavoidable. After all, there is no perfect celestial calibration
source. Therefore, the only practical approach is to conduct similar comprehensive
comparison studies using a variety of celestial sources and to identify systematic
differences commonly seen in various studies. More IACHEC papers will follow, which will
serve for this purpose.

The systematic differences should not be left as they are. We feel that the current
level of differences among various instruments (or even within the same instrument) can
be narrowed to a greater extent. For example, we described in this study that (1) the
Chandra ACIS-S3 full-frame and the sub-array data appear to have inconsistent
normalization, (2) the Swift XRT results remain inconsistent before and after the
substrate voltage changes or even at two epochs before the change for flux, (3) Suzaku
XIS is not well calibrated at the energies around the \ion{Si}{I} K edge, (4) the
EPIC-pn shows a deviation in the low energy tail of the spectrum, and (5) the three EPIC
instruments are inconsistent with each other.

Scientific measurements can be no better than the calibration accuracies. Continuous
calibration efforts are requisites to keep the instruments reliable and to maximize the
scientific output. We hope that the present study will help such efforts, which the
entire community relies upon.

\begin{acknowledgements}
We thank the IACHEC members for sharing their results prior to publication.  We
acknowledge Dai Takei and Kei Saitou for their help in Suzaku data reduction.
APB and AMR acknowledges support from the STFC.
\end{acknowledgements}

\bibliographystyle{aa}
\bibliography{ms}

\end{document}